\documentclass[letterpaper]{ar-1col}
\usepackage[utf8]{inputenc}
\usepackage{acronym}
\usepackage[numbers]{natbib}
\usepackage{amsmath,amssymb}
\usepackage{hyperref}
\usepackage{graphicx}
\usepackage{geometry}
\usepackage{color}
\usepackage[dvipsnames]{xcolor}
\usepackage[normalem]{ulem}

\usepackage{aas_macros}

\jname{Xxxx. Xxx. Xxx. Xxx.}
\jvol{AA}
\jyear{YYYY}
\doi{10.1146/((please add article doi))}

\newcommand{\be}{\begin{equation}}
\newcommand{\ee}{\end{equation}}
\newcommand{\bea}{\begin{eqnarray}}
\newcommand{\eea}{\end{eqnarray}}
\newcommand{\bel}{\begin{align}}
\newcommand{\eel}{\end{align}}

\def\l{\ell}

\def\M{{\cal M}}

\def\O{{\cal O}}

\def\Msun{\rm M_{\odot}}

\def\MTOV{M_{\max}^{\rm TOV}}
\def\MRNS{M_{\max}^{\rm RNS}}
\def\dd{\mathrm{d}}

\newcommand{\ie}{\textit{i.e.}}
\newcommand{\eg}{\textit{e.g.}}

\begin{document}

\markboth{Radice et al.}{Neutron Star Merger Dynamics}

\title{The Dynamics of Binary Neutron Star Mergers and of GW170817}

\author{David Radice$^{1,2,3}$, Sebastiano Bernuzzi$^{4}$, and Albino Perego$^{5,6}$
\affil{$^1$Institute for Gravitation \& the Cosmos, The Pennsylvania State University, University Park, PA 16802; email: david.radice@psu.edu}
\affil{$^2$Department of Physics, The Pennsylvania State University, University Park, PA 16802}
\affil{$^3$Department of Astronomy \& Astrophysics, The Pennsylvania State University, University Park, PA 16802}
\affil{$^4$Theoretisch-Physikalisches Institut, Friedrich-Schiller-Universit{\"a}t Jena, 07743, Jena, Germany}
\affil{$^5$Dipartimento di Fisica, Universit\'a di Trento, Via Sommarive 14, 38123 Trento, Italy}
\affil{$^6$INFN-TIFPA, Trento Institute for Fundamental Physics and Applications, Via Sommarive 14, I-38123 Trento, Italy}
}

\begin{abstract}
With the first observation of a binary neutron star merger through gravitational waves and light GW170817, compact binary mergers have now taken the center stage in nuclear astrophysics. They are thought to be one of the main astrophysical sites of production of r-process elements, and merger observations have become a fundamental tool to constrain the properties of matter. Here, we review our current understanding of the dynamics of neutron star mergers, in general, and of GW170817 in particular. We discuss the physical processes governing the inspiral, merger, and postmerger evolution, and we highlight the connections between these processes, the dynamics, and the multimessenger observables. Finally, we discuss open questions and issues in the field and the need to address them through a combination of better theoretical models and new observations.
\end{abstract}

\maketitle

\tableofcontents

\section{Introduction}
Neutron star (NS)\acused{NS} mergers are at the heart of some of the most pressing problems in nuclear astrophysics. Binary systems composed of two \acp{NS} (BNS) \acused{BNS} have provided the first evidence for the existence of \acp{GW}. 
The detection of a \ac{BNS} merger by LIGO/Virgo and \ac{EM} observer partners, GW170817, had a profound impact on our understanding of gravity, the physics of dense matter, the origin of \acp{SGRB}, and the site of production of r-process elements \cite{TheLIGOScientific:2017qsa, Abbott:2018wiz, GBM:2017lvd}. 
Many more multimessenger observations of \ac{NS} mergers are expected in the next years as the ground-based laser-interferometer detectors LIGO and Virgo reach their design sensitivity, and as KAGRA and LIGO India join the network \cite{Aasi:2013wya}.

\ac{GW} observation of inspiraling \acp{NS} can be used to measure the tidal deformability of the stars, probing the interior structure of \acp{NS}, and constrain the nature of matter at supernuclear densities \cite{Hinderer:2009ca, Damour:2012yf, DelPozzo:2013ala}. 
With third generation detectors, or for rare very nearby events, it will be possible to observe \ac{GW} emitted by the merger product of two \acp{NS} possibly constraining the presence of phase transitions at several times nuclear densities and temperatures of tens of MeV \cite{Sekiguchi:2011mc, Radice:2016rys, Most:2018eaw, Bauswein:2018bma}.

\ac{NS} mergers (NSNS and NS-black hole mergers) are also thought to be an important, if not a dominant, astrophysical site of production of r-process elements, such as gold \cite{Cowan:2019pkx}. The fact that \ac{NS} mergers produce some r-process nuclei is now firmly established by the multimessenger observations of GW170817 \cite{Cowan:2019pkx}. 
However, it is not clear whether \ac{NS} mergers produce all the r-process nuclei or if other astrophysical phenomena are required to explain the observed chemical abundances in our galaxy and satellites.

Isolated \acp{NS} are characterized by strong, but stationary gravitational fields. Their self-gravity (or compactness $C_A=GM_A/c^2R_A\sim0.15$, $A$ labeling one of the \acp{NS}) cannot be neglected. \ac{BNS} systems lose orbital angular momentum due to the emission of \acp{GW}, so \ac{BNS} spacetimes are dynamical. Nevertheless, the evolution of close circularized binaries can still be considered as an adiabatic process, as long as the radiation reaction timescale is much longer than the orbital period. In particular, the inspiral can be well described by a sequence of circular orbits until shortly before merger. As the two \acp{NS} approach each other, finite size (tides) and hydrodynamics effects becomes progressively more relevant and the inspiral terminates when the binary reaches the mass-shedding limit (Roche lobe overflow) \cite{Bejger:2004zx}.

Simulations in \ac{NR} are the most appropriate tool to study the dynamical phases of \ac{BNS} mergers: the late inspiral (the last ${\sim}20$ orbits), the merger, and its aftermath. Sophisticated models are required to quantitatively study all the features related to the merger and post-merger phase. State of the art simulations include dynamically evolving spacetime; finite temperature, composition dependent nuclear \ac{EOS}; \ac{GRMHD}\acused{GR}\acused{MHD}; weak interactions, and neutrino transport, although with different levels of approximations, \eg, Refs.~\cite{Sekiguchi:2011zd, Wanajo:2014wha, Foucart:2015gaa, Palenzuela:2015dqa, Sekiguchi:2016bjd, Kiuchi:2017zzg, Radice:2017zta, Fujibayashi:2017puw}. 

Here, we discuss the dynamics of \ac{BNS} mergers: their qualitative and quantitative features, the physics that controls the evolution of the binary, and the multimessenger signatures of the dynamics. Particular emphasis is on the nuclear astrophysics implications of mergers and on the comparison between theoretical predictions and GW170817, as well as on the new questions raised by the first detection. For a more general overview on \ac{NS} mergers we refer to Ref.~\cite{2016nure.book.....S}. We refer to Refs.~\cite{Kumar:2014upa, Fernandez:2015use, Metzger:2019zeh} for a more detailed discussion on the \ac{EM} emissions from \ac{BNS} mergers. The recent review by Shibata \& Hotokezaka~\cite{Shibata:2019wef} discusses the mass ejection from \ac{NS} merger in detail and is complementary to ours. We focus on the connection between outflow properties and specific physical processes and features of the postmerger dynamics.

The rest of this paper is organized as follows. Section \ref{sec:inspiral} discusses the inspiral phase with emphasis on the physics the two-body problem in \ac{GR} and on tidal effects. Section \ref{sec:merger} discusses the merger and postmerger evolution. We introduce physical processes operating in these phases and the associated timescales, and we discuss the outcome of mergers and of GW170817. Section \ref{sec:ejecta} is dedicated to the discussion of mass ejection from \ac{BNS} mergers and on how the features of the outflow depend on the postmerger dynamics. Finally, Section \ref{sec:conclusions} contains a summary of key points and of open questions in the field.

\section{Neutron Stars Inspiral}
\label{sec:inspiral}

If the binary forms through standard formation channels \cite{Aasi:2013wya}, eccentricity is efficiently radiated during the early evolution and by the time the binary enters the frequency band of ground-based GW interferometers the motion is circularized. 
Then the stars inspiral towards each other for the last few minutes, or thousands of orbits, of evolution emitting a \ac{GW} signal increasing in amplitude and frequency (chirp) until it reaches a maximum, conventionally denoted as \emph{moment of merger}.

\subsection{Two Body Dynamics}
The quasi-circular and quasi-adiabatic inspiral motion can be described within the \ac{PN} approximation to general relativity \cite{Blanchet:2013haa}. 
The \ac{PN} approximation applies to strongly self-gravitating compact binaries when the bodies are well-separated and the orbital angular velocity $\Omega$ is small, because it is an expansion in the relative velocity $v/c$ which is formally valid only if $v/c \ll 1$. The motion is characterized by an adiabaticity parameter $\dot{\Omega}/\Omega^2\ll 1$ that expresses the fact that radiation-reaction timescales is longer then the orbital timescale. 

The predicted gravitational signal is, at leading order, emitted at a frequency twice the orbital frequency with amplitude and phase scaling given by (quadrupole formula) 
\be
h(t) \sim \frac{1}{d} \M_c^{5/3} f^{2/3}_\text{GW}= \nu \frac{M}{d}(Mf_\text{GW}(t))^{2/3} \ , \ \ 
\phi(t)\sim 2 \M_c^{-5/8}t^{5/8}=2\nu^{-3/8}(t/M)^{5/8} \ ,
\ee
where $\M_c=M\nu^{3/5}$ is the chirp mass, $M=M_A+M_B$ is the binary mass, $\nu=M_A M_B/M^2$ is the symmetric mass ratio, $f_\text{GW}=\dot{\phi}$, $d$ is the source distance, and, for clarity, we have suppressed geometric factors and assumed $G = c = 1$. Note the problem trivially scales with the total mass $M$ of the system as long as the bodies can be considered point-masses (this holds in general, also for binary black holes). Tidal interactions due to the finite size of the bodies are effects that formally enter the action at fifth \ac{PN} order. As discussed below, the \ac{BNS} dynamics is influenced by the tidal interactions of the two stars at the frequencies relevant for ground-based detector observations. 

The \ac{PN} approach describes qualitatively the inspiral phase but, since \ac{PN} is an asymptotic expansion, it ultimately fails to quantitatively describe binaries in the high-frequency regime, \ie, $f_{\rm GW}\gtrsim 50$~Hz. The \ac{EOB} is a formalism to solve the GR two-body problem that can be applied to both the low- and the high-velocity regimes \cite{Buonanno:1998gg}. 
The \ac{EOB} is a relativistic generalization of the well-known Newtonian property that the relative motion is equivalent to the motion of a particle of mass $\mu= \nu M$ in an effective potential. The GR dynamics can, in fact, be mapped into the motion of an effective particle $\mu$ into an effective metric. The \ac{EOB} is an Hamiltonian formalism describing semi-analytically the inspiral-merger-ringdown dynamics of binary black holes; Damour \& Nagar \cite{Damour:2009wj} incorporated the treatment of tidal effect into the formalism, thus extending the model's applicability to BNS. We refer to \cite{Damour:2012mv} for a review, but recall here that \ac{EOB} is a unified framework to incorporate different perturbation approaches to the two-body problem, resum the \ac{PN} series, and include nonperturbative information from simulations.

\subsection{Tidal Effects}
The description of tidal interactions in the \ac{PN} dynamics of self-gravitating and deformable bodies was formulated in a series of works by Damour, Soffel and Xu in the '90s \cite{1983grr..proc...58D}.

They developed a multi-chart approach whereby an {\it outer problem}, in which the bodies are ``skeletonized'' as worldlines with global properties, is matched to an {\it inner problem}, in which the effects of the other bodies in the worldtube around a given body are included. In the case of compact binaries, the inner problem corresponds to the description of the tidal response of a NS due to the external gravitational field of the companion. The matching with the outer problem allows one to include the effect of the tidal deformations on the orbital dynamics and the \ac{GW} radiation. The presentation here follows closely that of Refs.~\cite{Damour:2009vw, Damour:2009wj}. 

A fully relativistic treatment of the \emph{inner problem} was developed in \cite{Hinderer:2007mb, Damour:2009vw, Binnington:2009bb}. In the local frame of body $A$, the internally-generated mass $M^A_L$ and spin $S^A_L$ multipole moments, $L=i_1i_2...i_\l$ being a multi-index, are related to the external gravitoelectric $G^A_L$ and gravitomagnetic $H^A_L$ tidal moments\footnote{The tidal moments are defined as the symmetric-trace-free projection of the derivatives of the externally-generated parts of the local gravitoelectric $\bar{E}_a$ and gravitomagnetic fields $\bar{B}_a$, e.g. $G^A_L=\partial_{\langle L-1}\bar{E}^A_{a_\l \rangle}|_{X^a\to0}$, where $X^a$ are local coordinates \cite{Damour:2009vw}.} by the {\it tidal polarizability coefficients}, 
\be
M^A_L = \mu_\l G_L^A \ \ , \ \ S^A_L = \sigma_\l H_L^A \ .
\ee
The gravitoelectric (gravitomagnetic) coefficient $G\mu_l$ has dimension $[{\rm length}]^{2\l+1}$ and measures the $\l$-th-order mass (spin) multipolar moment induced in the NS by the external $\l$-th-order gravitoelectric (gravitomagnetic) field. The dimensionless relativistic Love numbers are defined as
\be \label{eq:Lovenum}
k_\l = \frac{(2\l-1)!!}{2}\frac{G\mu_\l}{R^{2\l+1}} \ \ , \ 
j_\l = \frac{(2\l-1)!!}{2}\frac{G\sigma_\l}{R^{2\l+1}} \ ,
\ee
with $R$ being the NS radius. Note that the many works in the literature focus on the dominant quadrupole $\l=2$ gravitoeletric coefficient and drop the subscript, \eg, Ref.~\cite{Hinderer:2007mb}. For \acp{BH} $\mu_\l^{\rm BH}=\sigma_\l^{\rm BH}=0$ \cite{Damour:2009vw, Binnington:2009bb}. 

In practice, the calculation of the Love numbers reduces to the solution of stationary perturbations of spherical relativistic stars, because it is assumed that the external field varies sufficiently slowly (``adiabatic tides''). The tidal coefficients have a strong dependency on the NS compactness. Thus, Love numbers must be computed in \ac{GR}, and not in the Newtonian limit. Love numbers depend on the \ac{EOS} employed to construct the equilibrium \ac{NS}. Hence, they carry the imprint of the \ac{EOS} on the binary dynamics.

If the external field is dynamical, the star response can be described, at linear order in the deformation, as a superposition of the star's proper modes. Modes are excited when the orbital frequency matches their resonant frequency. The problem has been studied extensively in Newtonian gravity,

in \ac{GR} for a test-mass orbiting a \ac{NS}, and for comparable masses in \ac{PN} theory \eg, \cite{Ho:1998hq, Pons:2001xs, Steinhoff:2016rfi}. Such ``dynamical tides'' are dominated by the fundamental pressure modes ($f$-modes), but typically in a non-resonant way, since the $f$-modes resonance is in the ${\sim}$kHz regime, which would correspond to the merger and postmerger phases, past the point where the stars exist as separate objects. Resonances can be excited for other types of modes, \eg, $g$-modes or $r$-modes, because these have lower frequencies, but their energies are smaller.

Finite-size effects are incorporated into the \ac{PN} two-body dynamics by augmenting the effective action  
\be
S = S_{\rm GR} + S_{\rm pointmass}
= \frac{1}{16\pi G}\int R \sqrt{|g|} \dd x - \sum_A \int M_A \dd s_A \ ,
\ee
the second term being the skeletonized description as point masses, with the nonminimal (worldline) couplings 
\be
S_{\rm nonminimial}  = \sum_A 
\frac{\mu_\l^A}{2\l!} \int (G^A_L)^2 \dd s_A +
\frac{\l\, \sigma_\l^A}{\l! 2(\l+1)} \int (H^A_L)^2 \dd s_A \ .
\ee
The additional term alter the dynamics at 5PN in a way that is linear in the tidal deformations. The tidal contribution to the two-body Lagrangian at leading \ac{PN} (Newtonian) order contains only the $\l=2$ gravitoelectric terms and reads 
\be \label{eq:Ltidal}
L_{\rm tidal}^\text{LO} = 
k^A_2 GM_B^2\frac{R^{5}_A}{r^{6}} + 
(A\leftrightarrow B)
\ ,
\ee
where $r$ is the separation between the stars in the binary. Eq.~\eqref{eq:Ltidal} indicates that tidal corrections are attractive and short range. The effect of tides can be illustrated considering the modification to the Kepler law given by the quadrupolar gravitoelectric term
\be
\Omega^2 r^3 = GM\left[ 1+ 12\frac{M_A}{M_B}\frac{R_A^5}{r^5}k^A_2 +
  (A\leftrightarrow B)\right] \ .
\ee 
At a given radius the frequency is higher if the tidal interactions are present. In other words, the motion is accelerated by tidal effects and the system merges earlier and at a lower frequency. The contact \ac{GW} frequency of the two NS can be estimated setting $r=R_A+R_B$ and finding $2GM\Omega\simeq 2(M_B/(M C_B) + M_B/(M C_B))^{-3/2}$ \cite{Damour:2009wj}. For equal masses the latter relation translates to
\be
f_\text{GW}^\text{contact} \simeq 1.327\, \left(\frac{C}{0.15}\right)^{3/2} \left(\frac{M}{2.8\Msun}\right)\, {\rm kHz} \ .
\label{eq: GW peak frequency}
\ee
Simulations show that the contact between the two \acp{NS} happens approximately $2{-}4$ \ac{GW} cycles prior to merger at an even lower frequency $f_{\rm GW}^{\rm NR, contact} \simeq 700\ (M/2.8\Msun)\ {\rm Hz}$ \cite{Bernuzzi:2012ci}.

The EOB two-body Hamiltonian for nonspinning binaries is written in terms of an effective Hamiltonian: 
\be
H_{\rm EOB} = M\sqrt{1+2\nu(\hat{H_{\rm eff}}-1)} \ ,
\ \ 
\hat{H}_{\rm eff} = \frac{H_{\rm eff}}{\mu} 
= \sqrt{A(u;\nu)(1+p^2_\phi u^2+2\nu(4-3\nu)u^2p^4_{r*}) +p^2_{r*}} \ .
\ee
where $u=GM/rc^2$ is the Newtonian potential. The effective Hamiltonian $H_{\rm eff}$ reduces to the Hamiltonian of a particle in Schwarszchild spacetime for $\nu\to0$ where $A(u;0)=1-2u$. For finite mass ratio the function $A(u;\nu)$ is computed from \ac{PN} results, and it is completely known analytically up to 4PN \cite{Damour:2015isa} 

for the point-particle dynamics, only some of the point-particle terms are known at 5PN. The expression at 4PN is remarkably simple
\be
A_0 = 1- 2u + \nu ( 2 u^3 + a_4 u^4 + a_5(\nu,\ln u) u^5 ) \ ,
\ee
where $a_4=(94/3 - 41/32\pi)$, and $a_5(\nu,\ln u)$ is a linear function of $\nu$ and $\ln u$. For BBH applications, the function $A_0$ is further resummed using analytical techniques (\eg, Pade functions), and the 5PN parameters that are not known analytically are fixed using \ac{NR} results. Tidal interactions are included by augmenting the potential $A = A_0 + A_{\rm tidal}$, with an expression inferred from the above tidal Lagrangian \cite{Bini:2012gu}. The tidal potential has the form
\be
A_{\rm tidal} = \sum_{\l\geq2} \left[
\kappa^{A+}_\l u^{2\l+2} (1+\alpha^{(\l+)}_1 u+\alpha^{(\l+)}_2 u^2+ ...) 
+\kappa^{A-}_\l u^{2\l+3} (1+\alpha^{(\l-)}_1 u+ ...) 
+ (A\leftrightarrow B)
\right]
\ee
where $\alpha^{(\l)}_i(\nu)$ are coefficients and 
\be
\kappa^{A+}_\l = 2 k_\l^A\left(\frac{M_A}{M C_A}\right)^{2\l+1}\frac{M_B}{M_A} \ , \ \ 
\kappa^{A-}_\l = 2 j_\l^A\left(\frac{M_A}{M C_A}\right)^{2\l+1}\frac{M_B}{M_A} \ ,
\ee
are multipolar tidal polarizability coupling constants. The current analytical knowledge comprise gravitoelectric terms $\ell=2,3$ up to the next to next leading order (NNLO; coefficients $\alpha^{(2+,3+)}_{1,2}$) and gravitomagnetic terms up to NLO (coefficient $\alpha^{(2-)}_{1}$).

Taking the Newtonian limit illustrates the meaning of the above formulas
\be
H_{\rm EOB} \simeq M c^2 + \frac{\mu}{2} p^2 + \frac{\mu}{2}(A-1) 
= M c^2 + \frac{\mu}{2} p^2 + \frac{\mu}{2} \left( -\frac{2GM}{c^2 r^2} + ... 
- \frac{\kappa_2^T}{r^5} \right) \ .
\ee
The constant $\kappa_2^T=\kappa^A_2+\kappa^B_2$ encodes the effect of tidal interactions at leading order. For a large span of EOS, masses in $[1,2]\Msun$ and mass ratios in $q\in[1,2]$ its values are $\kappa^T_2\sim[50,500]$. A common alternative notation (but more cumbersome) uses the quantities $\Lambda_2^i\equiv 2/3 k^i_2 (c^2 R_i/GM_i)^{5}$ with $i \in \{A,B\}$, in place of the $\kappa^A_2$ and defines 
\begin{align}
  \label{eq:LambdaT}
  \tilde\Lambda &= \frac{16}{13}
  \frac{(M_\text{A} + 12 M_\text{B}) M_\text{A}^4}{M^5}\Lambda_\text{A} + (A\leftrightarrow B)\ .
\end{align}

The conservative dynamics described above is complemented by a waveform providing the radiation reaction for the dynamics and the emitted radiation \cite{Damour:2008gu}. Tidal corrections are introduced also in the waveform 

\cite{Damour:2012yf, Banihashemi:2018xfb}. 
At leading order the stationary phase approximation of the waveform reads
\be
h(f) = A f^{-7/6} e^{-i(\Psi_0(x)+\Psi_{\rm tidal}(x))} =
A f^{-7/6} e^{-i(\Psi_0(x) - 39/4 \kappa^{T}_2 x^{5/2})} \ ,
\ee
where $x(f)=(\pi GM f/c^3)^{2/3}$ and $\Psi_0(x)$ is the point-mass phase. Note that the tidal contribution at leading order is again fully determined by $\kappa^T_2$. For this reason, the latter (or equivalently $\tilde\Lambda$) is the quantity that is best measured from \ac{GW} observations. 

The validity of the \ac{EOB} description of tides has been tested in the high-frequency regime against long-term \ac{NR} simulations starting at about $500$~Hz and lasting about 10-20 orbits up to merger \cite{Baiotti:2011am, Bernuzzi:2012ci, Bernuzzi:2014owa, Dietrich:2017feu, Akcay:2018yyh}. Using the \ac{EOB} point-mass dynamics as a baseline, the \ac{PN} expression for $A_{\rm tidal}$ reproduces remarkably well the \ac{NR} results to within their estimated errors, but it becomes inaccurate in the very last orbits, or for large values of $\kappa^T_2$. Advanced tidal EOB models have been proposed in \cite{Bernuzzi:2014owa, Nagar:2018zoe, Akcay:2018yyh, Nagar:2018plt} using high-order results from gravitational self-force calculations of tides (TEOBResumS) and in \cite{Hinderer:2016eia} implementing dynamical tides (SEOBNRT). These models currently reproduce a large sample of the available \ac{NR} waveforms within the numerical uncertainties. 
In presence of spins new tidal contributions arise \cite{Poisson:1997ha, Pani:2015hfa}. For example, a rotating star's oblateness creates a deformation in the gravitational field outside the star, which is measured by the quadrupole tensor. This effect, quadratic in the star's self-spin, generates an attractive contribution to the potential that affect the inspiral motion at second PN order $\O (v/c)^4$ \cite{Poisson:1997ha}. 
Finally, we mention that phenomenological tidal models fitting hybrids EOB and NR data with simple formulas are used for efficient GW analysis \cite{Dietrich:2017aum, Kawaguchi:2018gvj}.

\subsection{Gravitational Waves}
\begin{figure}
    \centering
    \includegraphics[width=\textwidth]{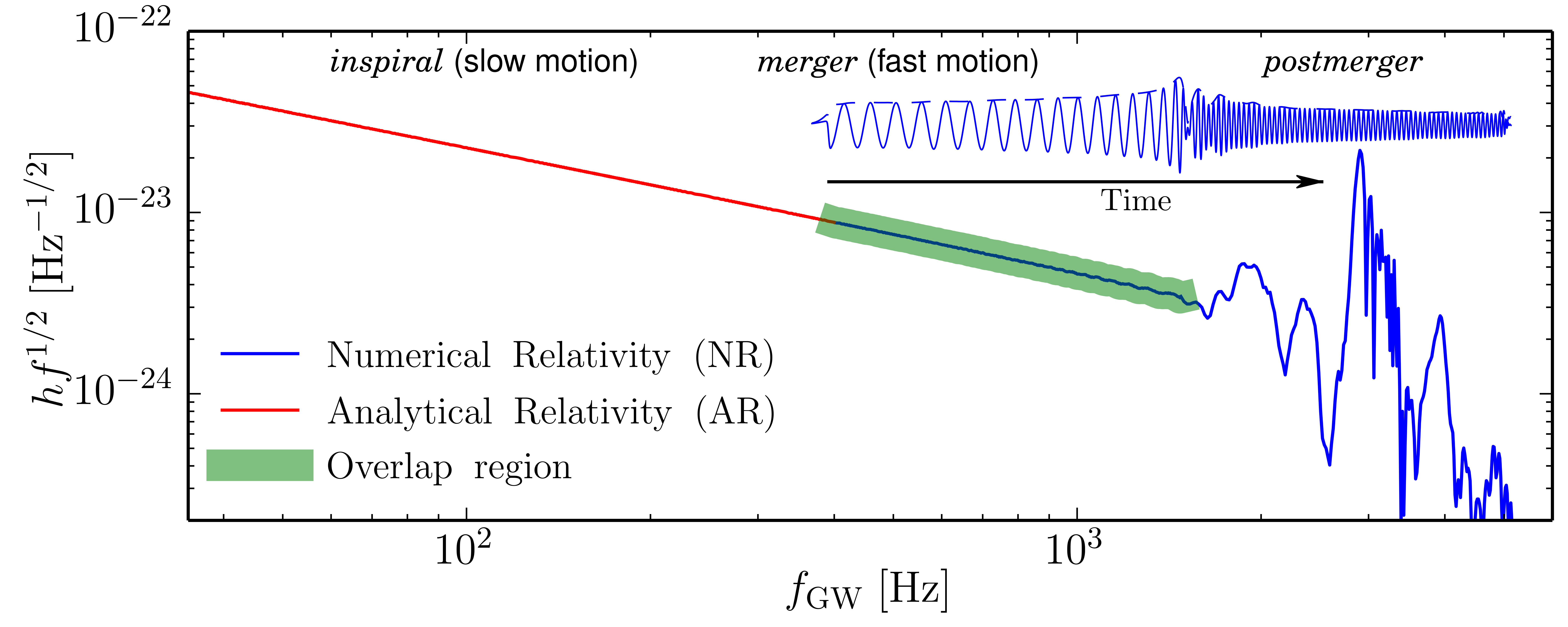}
    \caption{Schematic representation of the complete GW signal from BNS mergers. The inspiral phase can be well described using the tools of analytical relativity, while the postmerger phase can only be described with numerical relativity. Complete waveform models are constructed by matching the two approaches in the region where both are valid.}
    \label{fig:gwsignal}
\end{figure}

The formalism described above delivers an accurate GR prediction of the \ac{BNS} waveform in the complete frequency range of ground-based interferometers, ${\sim}10{-}2048$~Hz. \ac{EOB} models are the only ones able to predict the \ac{GW} waveform in the frequency region where \ac{PN} theory breaks down and \ac{NR} simulations are neither available nor feasible. Systematic differences between \ac{PN} approximants at different orders are present already at \ac{GW} frequencies as low as 50~Hz \cite{Damour:2012yf} and can impact the GW parameter estimation at sufficiently high \ac{SNR}. While long-term NR simulations are possible, controlling the phase errors within the sub-radiant precision over tens-to-hundreds of orbits remains an open challenge \cite{Bernuzzi:2016pie}. 
\ac{EOB} models can be completed in the kiloHertz regime with models describing the high-frequency emission from the merger remnant \cite{Breschi:2019srl}. The latter can be inferred only from NR hydrodynamics simulations and usually model the characteristic early-time burst signal, \eg, \cite{Bernuzzi:2015rla, Chatziioannou:2017ixj, Easter:2018pqy}. 
This is summarized in Fig.~\ref{fig:gwsignal}, which shows the full \ac{GW} spectrum and for a \ac{BNS} system, as well as the time domain waveform in the last few cycles prior to and after the merger. 

We remark that tidal interactions play a key role in determining the late-inspiral and merger dynamics. For example, a binary with two NS with comparable masses ${\sim}1.4\Msun$ and low spins performs about ${\sim}1300$ revolutions from $30$~Hz and merges at ${\sim}1.5{-}2$~kHz (depending on EOS). In absence of tidal interactions the same binary would merge at a frequency three times larger and accumulate a dephasing of one radiant at ${\sim}200$~Hz and about ten radians up to the merger frequency of the BNS.

\begin{figure}
    \centering
    \includegraphics[width=0.49\textwidth]{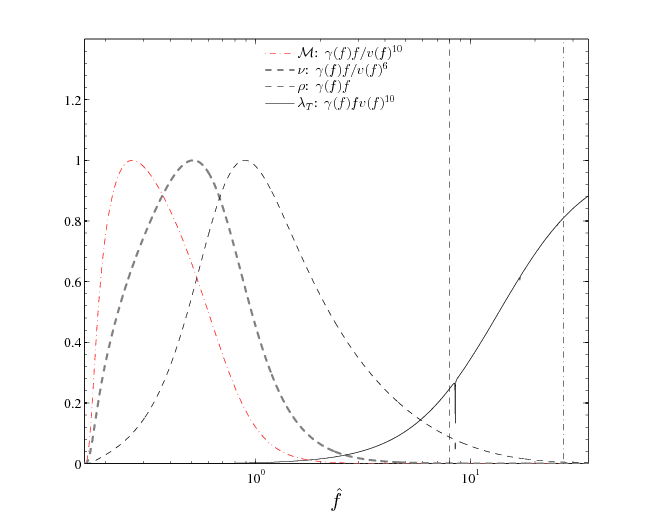}
    \caption{Measurability of the binary chirp mass $\mathcal{M}$, symmetric mass ratio $\nu$, and tidal deformability parameter $\lambda_T$ ($=32 M^5 \tilde\Lambda$ in our notation) as a function of the normalized \ac{GW} frequency $\hat{f} = f/f_0$, $f_0 \simeq 57\ {\rm Hz}$ being the frequency at which the SNR density $\rho$ is maximum. The analysis consider a LIGO at design sensitivity (\texttt{ZERO\_DET\_high\_P} configuration) and $M_A = M_B = 1.4\ \Msun$. The rightmost vertical line denotes the contact frequency (Eq.~\ref{eq: GW peak frequency}), while the other dashed line denotes $450\ {\rm Hz}$. Reproduced from Damour, Nagar, and Villain, \textit{Measurability of the Tidal Polarizability of Neutron Stars in Late-Inspiral Gravitational-Wave Signals}, Phys. Rev. D85 (2012) 123007, Ref.~\cite{Damour:2012yf}. Copyright the APS. Reproduced with permission.}
    \label{fig:damour.2012}
\end{figure}

Under the assumption of Gaussian noise and high \ac{SNR}, and considering the PN waveform, the measurability of a given parameter is determined from the diagonal elements of the Fisher matrix by integrals of type $I_p=\int \dd \ln f \gamma(f) f x^{2p}(f)$, where $\gamma(f)df$ is a measure that depends on the noise of the detector \cite{Cutler:1994ys, Damour:2012yf}. Hence, the distribution of information on the \ac{GW} frequency range is mainly determined by the integrand of $I_{p}$; see Ref.~\cite{Damour:2012yf} and Fig.~\ref{fig:damour.2012}. For a fiducial BNS, the chirp mass (related to $I_{-10}$) is almost entirely determined by the signal at low-frequencies ${\lesssim}30$~Hz. Information on the symmetric mass ratio ($I_{-6}$) and the \ac{SNR} ($I_0$) is also primarily given by the useful GW cycles below 50 and 100 Hz respectively. By contrast, the measurability of tidal parameters is related to $I_{+10}$. Thus, while the total mass can be extracted rather accurately with nontidal templates, capturing the GW phasing above 100 Hz requires tides.

A straightforward argument based on the Newtonian equations presented above indicate that the merger dynamics is primarily determined by  $\kappa^T_2$ \cite{Bernuzzi:2014owa}. This expectation has been directly verified to the percent level with more than hundreds \ac{NR} simulations \cite{Zappa:2017xba, Breschi:2019srl}. For example, the \ac{GW} frequency at the time of merger can be fitted to the percent level with

\be
\label{eq:fmrg.nr}
f^\text{merger}_\text{GW} \simeq 2.405 \left(\frac{1 + 1.307\, \cdot 10^{-3} \xi}{1 + 5.001\,\cdot 10^{-3} \xi }\right) \left(\frac{M}{2.8 \Msun}\right)\ {\rm kHz} \ ,
\ee
where 
$\xi = \kappa_2^T + 3200 (1-4 \nu)$
. Similar relations exists for all the relevant dynamical quantities, such as the binding energy, the angular momentum, or the GW luminosity at merger \cite{Bernuzzi:2014owa, Zappa:2017xba, Breschi:2019srl}. These relations are often called ``quasi-universal'' or ``EOS-insensitive'' because once the quantities are appropriately rescaled by the binary mass and symmetric mass ratio, they are simple functions of the the mass ratio and of $\kappa^T_2$, the latter of which encodes all the EOS information. Note that, even though the errors introduced by the \ac{EOB} approach are maximum at merger (moment at which the description of the system as a binary breaks down), the \ac{EOB} results still agree to within ${\lesssim}20$\% with the \ac{NR} fitting formulae discussed above.

In the case of GW170817 most of the \ac{SNR} was accumulate in the frequency range 30 Hz to 600 Hz, roughly corresponding to the last 1300 orbits to merge for an equal-mass binary with total mass $M\simeq 2.7\Msun$. GW170817 is compatible with a BNS system with chirp mass $\mathcal{M} = 1.186(1)\Msun$, mass ratio $q \in [1,1.34]$ and $\tilde{\Lambda}\simeq 300$ and smaller than ${\sim}800$ \cite{TheLIGOScientific:2017qsa, Abbott:2018wiz, LIGOScientific:2018mvr}. The constraint on $\tilde\Lambda$ translates to $\kappa^T_2\lesssim 150$. Among the different waveform approximants used in the analysis \cite{LIGOScientific:2018mvr}, EOB models favors slightly larger median values for $\tilde\Lambda$ (larger radii) than the others; but all results are compatible at the 90\% confidence level. Also, if priors include a lower bound on $\tilde\Lambda$ inferred from the interpretation of the the EM counterpart, then larger values of $\tilde\Lambda$ are favoured \cite{Radice:2017lry, Radice:2018ozg}.

The mass ratio and the individual masses for GW170817 are less precisely determined, and there are systematic uncertainties also related to the spin priors \cite{Abbott:2018wiz}. Since the tidal parameters are partially degenerate with the mass ratio, these uncertainties affect also the \ac{EOS} constraints derived from GW170817. When low spin priors (dimensionless NS spins assumed to be $\lesssim0.05$) are assumed, the individual radii of the NS are inferred to be about $R\sim 11-12$~km \cite{De:2018uhw, Abbott:2018exr}, where the most precise measurement at 90\% credible level $R\simeq 11.9\pm1.4$~km is obtained with the additional requirement that the \ac{EOS} must support nonrotating \acp{NS} with masses of at least $1.97\Msun$ \cite{Abbott:2018exr}.

The inspiral signal and tidal phasing can directly constrain regions in the EOS pressure-density diagram \cite{Abbott:2018exr}. The pressure is best constrained at around the maximum density of the NSs in the binary \cite{Agathos:2019sah} that for the fiducial BNS is $\rho_{\max}\simeq 2\rho_{0}$. Ref.~\cite{Abbott:2018exr} finds $P(2\rho_0)=3.5^{+2.7}_{-1.7} \times10^{34}\ {\rm dyn}\ {\rm cm}^{-2}$ at the 90\% level.

The merger GW signal was not observed, but the GW frequency at merger can be accurately predicted from the probability distribution of $\tilde{\Lambda}$ using the NR fit (Eq.~\ref{eq:fmrg.nr}). One finds that it falls in the range $1.2{-}2$~kHz \cite{Breschi:2019srl}. Similarly, the peak luminosity is estimated to be larger than $0.1\times10^{56}$ erg/s \cite{LIGOScientific:2018mvr}. The sensitivity of the detectors in August 2017 was insufficient to clearly identify a signal at frequencies $f\gtrsim f_\text{mrg}$ \cite{Abbott:2017dke, Abbott:2018hgk}, but if the merger had produced a NS remnant the main peak frequency of the postmerger signal should have been located at $2.5{-}3.2$~kHz \cite{Breschi:2019srl}.

\begin{figure}
    \centering
    \includegraphics[width=\textwidth]{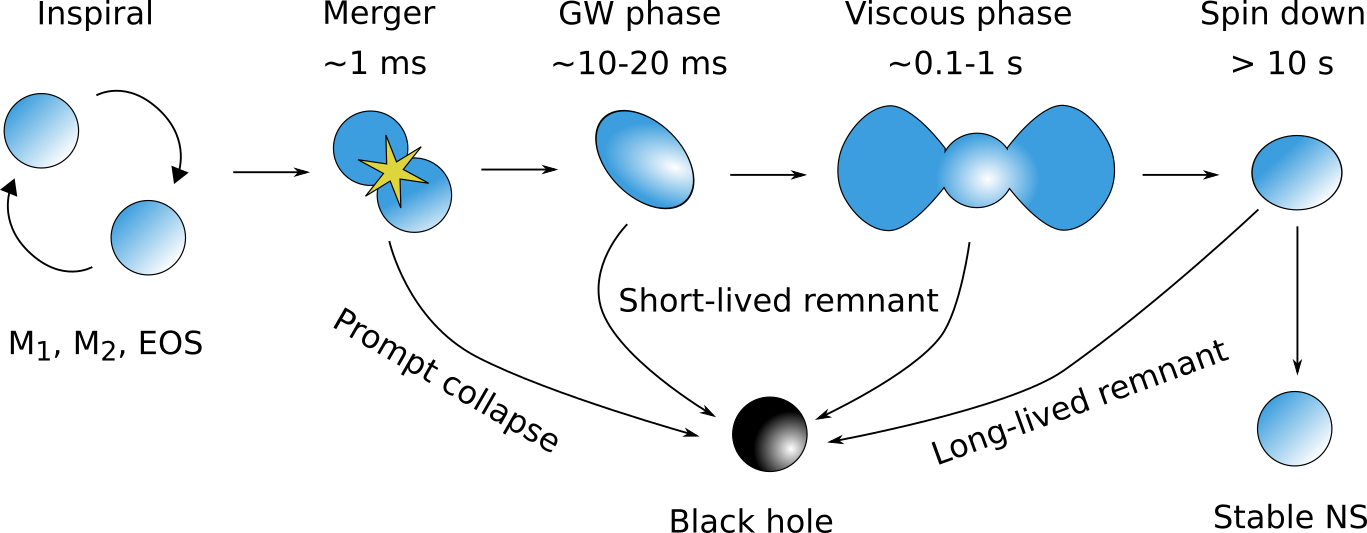}
    \caption{Overview of the different phases in a NS mergers and of the relative timescales. The inspiral ends with the merger, when the two stars start to fuse together. The early postmerger evolution is entirely driven by hydrodynamics and by GW emission. If the remnant does not collapse within ${\sim}10{-}20\ {\rm ms}$, GW losses subside and other physical processes become more important: angular momentum redistribution due to turbulent viscosity and neutrino losses operate over a timescale of a tenth of a second to a few seconds. This is also the characteristic timescale for the evolution of the remnant disk. If the remnant does not collapse over a timescale of a few seconds, then it will spin down due to MHD effects over a possibly much longer timescale of several seconds to a few hours.}
    \label{fig:overview}
\end{figure}

\section{Merger and Postmerger}
\label{sec:merger}
As the \acp{NS} come into contact and the inspiral terminates, the dynamics of the system becomes increasingly complex. Matter is compressed and heated up to extreme densities and temperatures, and new physical effects, such as \ac{MHD}\acused{GRMHD} turbulence and neutrino-matter interactions, become important (\S \ref{sec:dynamics.and.thermodynamics}) and can affect the outcome of the merger in ways that are not completely understood (\S \ref{sec:merger.outcome}). Before we discuss the related physics in details, we give in Fig.~\ref{fig:overview} a first overview of the \ac{BNS} dynamics after merger. \ac{BH} formation might be the immediate outcome of the merger, or it could be delayed by milliseconds to minutes. It is also not excluded that some \ac{BNS} might even form stable \ac{NS} remnants. After a first phase in which \ac{GW} emission and hydrodynamics play the most important role, \ac{GW} emission decays and angular momentum transport due to \ac{MHD} stresses and neutrino emission and re-absorption takes over. Over longer timescales, if the remnant has not yet collapsed to \ac{BH}, the system spins down due to residual \ac{GW} losses and \ac{EM} torques. 

\subsection{Dynamics and Thermodynamics Conditions}
\label{sec:dynamics.and.thermodynamics}

During the binary inspiral, the \ac{NS} matter is assumed to be in cold, neutrinoless, weak equilibrium, and degenerate baryons are the major source of pressure. 

Tidal deformation dissipates energy, but the increase in temperature $\Delta T \lesssim 0.1~{\rm MeV}$ and the neutrino losses are marginal up to the final phase of the inspiral \cite{Lai:1993di}. Thus, this equilibrium composition is maintained up to merger.

The binary orbital speed at merger can be estimated as $v_{\rm orb} \simeq \Omega\, r \simeq \sqrt{GM/(R_A+R_B)}$ and for an equal mass merger it reads
\be
v_{\rm orb}/c \simeq \sqrt{C} \simeq 0.39 \left( C/0.15 \right)^{1/2}\ .
\ee
Since during the inspiral the \ac{GW} frequency is approximately twice the orbital frequency and at leading order its evolution satisfies $\dot{\Omega}_{GW}^3 \sim (3456/125) (G \mathcal{M}_c/c^3)^{5} \Omega_{\rm GW}^{11}$, the radial infall velocity $v_r \simeq 2\, \Omega\, r\,  \dot{\Omega}/(3 \Omega^2)$ can be estimated as
\be
v_r/c\simeq \frac{192 \pi}{15} \frac{G^3 M^3}{c^5 \left( R_A + R_B \right)^3}\frac{q}{(1+q)^2}  \ . 
\ee
For an equal mass merger $v_r/c \simeq 0.034 \left( C/0.15 \right)^3$. Since $v_{\rm rad} \ll v_{\rm orb}$, the dynamics is primarily dominated by the orbital motion and 
\be
t_{\rm merger} \simeq 1/ \left( 2f^{\rm contact}_{\rm GW} \right) \simeq 1.50 \, {\rm ms}~\left( M/2.8~\Msun{} \right)^{-1/2} \left( C/0.15 \right)^{-3/2}
\ee
for NSs of comparable masses. Clearly, more massive binaries and more compact \ac{NS} result in faster and more violent mergers. Matter coming from each of the two \ac{NS} slip past each another at the contact interface and Kelvin-Helmholtz instability occurs. The two \ac{NS} cores, which initially reside behind this contact interface, fuse over a time scale of a few $t_{\rm merger}$.

\begin{figure}
    \begin{minipage}{0.49\textwidth}
        \includegraphics[width=\textwidth]{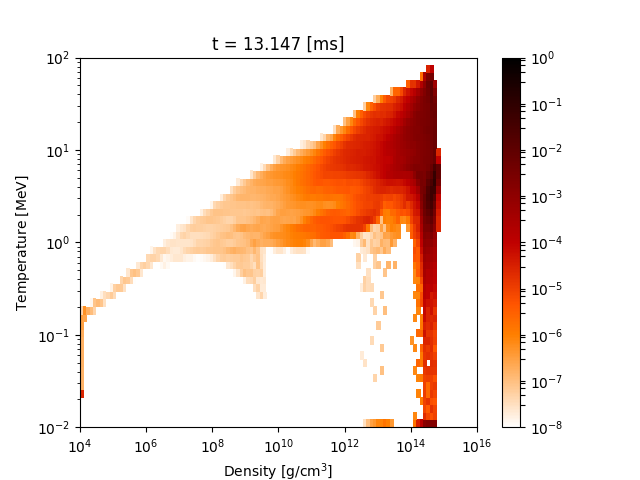}
    \end{minipage}
    \hfill
    \begin{minipage}{0.49\textwidth}
        \includegraphics[width=\textwidth]{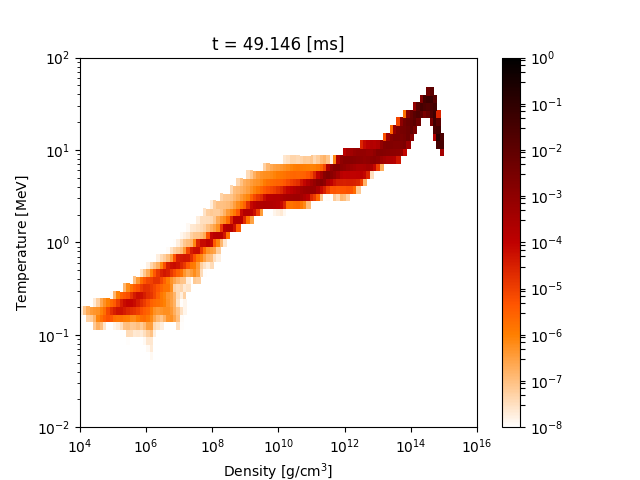}    
    \end{minipage}
    \caption{Histograms of the thermodynamics conditions experienced by matter during a \ac{BNS} merger from an equal mass merger simulation extending up to $40{\rm ms}$ postmerger and employing the DD2 \ac{EOS}. Left panel: Immediately after the merger, when the peak temperatures are reached, matter around and below nuclear saturation density reach temperatures of several tens of MeV. The \ac{NS} cores are visible as ultra-high density, low temperature regions, while the disk forming at lower densities is high non-homogeneous. Right panel: at much later times, the two cores have fused into a single one. The rotational dynamics, coupled with the adiabatic expansion, has driven the remnant towards axisymmetric thermodynamics conditions. Reproduced from Perego, Bernuzzi, and Radice, \textit{Thermodynamics Conditions of Matter in Neutron Star Mergers}, Eur.Phys.J. A55 (2019) no.8, 124, Ref.~\cite{Perego:2019adq} with kind permission of The European Physical Journal (EPJ). An animated version of this figure for a binary simulations employing the SFHo EOS is available as supplemental material. Follow the Supplemental Materials link in the online version of this article or at http://www.annualreviews.org/.}
    \label{fig:histograms}
\end{figure}

The forming remnant is initially far from hydrodynamical equilibrium: episodes of (gravity-driven) matter compression and (nuclear- and centrifugally-driven) expansion follow one another and the remnant bounces several times. The maximum density and temperature increase immediately after merger as a consequence of matter compression and oscillate due to the bounce dynamics \cite{Perego:2019adq}. Despite the large relative collision speed, the high speed of sound of matter at nuclear and supra-nuclear densities ($c_s \gtrsim 0.2c $ for $\rho \gtrsim \rho_0$) prevents the formation of hydrodynamical shocks inside the two coalescing cores. Only at the surface of the massive \ac{NS} pressure waves can steepen into shock waves which accelerate matter at the edge of the remnant up to mildly-relativistic speeds (\S \ref{sec:ejecta}). Thus, matter inside the cores remains cold ($T \lesssim 10\,{\rm MeV}$; $s \lesssim 1 {k_{\rm B}/{\rm baryon}}$) during the entire merger process. This is clearly visible in Fig.~\ref{fig:histograms}, where the thermodynamical conditions of matter during the merger are presented.

While the densest part of the cores rotate and fuse, compressed matter at the contact interface is pushed outwards. Compression and shear dissipation increase its temperature (up to $T\sim 70{-}110~{\rm MeV}$,
see Fig.~\ref{fig:histograms}) forming a pair of co-rotating hot spots displaced by an angle of ${\sim} \pi/2$ with respect to the densest cores \cite{Kastaun:2016yaf}. 

This structure survives until the cores have completed their fusion (or until BH formation). At that point the hot spots have evolved into a hot annulus. The core of the remnant remains relatively cold instead.

Material expelled from the central part of the remnant due to tidal torques or from the collision interface settle into a thick accretion disk, with typical aspect ratio $H/R \sim 1/3$ and mass between $0.001{-}0.2\ \Msun{}$. A phenomenological fit in terms of the $\tilde{\Lambda}$ parameter has been proposed and combined with the \ac{GW} data to derive a new constrain on $\tilde\Lambda$ in Refs.~\cite{Radice:2017lry, Radice:2018xqa, Radice:2018ozg}. Because of the different temperatures in the tidal tail (cold) and collisional interface (hot), the disk is initially highly non-uniform, as visible in the left panel of Fig.~\ref{fig:histograms}. As a consequence of the fast expansion, densities and temperatures drop inside the forming disk. Since the \ac{EOS} is dominated by non-relativistic baryons and the expansion proceeds mostly adiabatically, their evolution satisfies $T^3/ \rho^2 \sim {\rm const}$. 

During the core fusion phase, the remnant is characterized by a pronounced ($m=2$) bar deformation that powers a significant emission of GWs over the first ${\sim}10{-}20\ {\rm ms}$ after merger and launches spiral waves into the disk. The emission of energy and angular momentum provides a backreaction that quickly damps the bar mode. Thus, the GW emission is the major driver of the dynamics in the immediate aftermath of the merger. We refer to this phase as the GW-dominated postmerger phase.

The continue action of shocks and spiral waves increases the entropy in the disk and eventually produces an axisymmetric Keplerian disk characterized by a temperature profile that changes smoothly from $\sim$10~MeV (for $ \rho \simeq 10^{13}{\rm g~cm^{-3}}$) down to ${\sim}0.1$~MeV (for $ \rho \simeq 10^{4}{\rm g~cm^{-3}}$), as visible in the right panel of Fig.~\ref{fig:histograms}. Correspondingly, the entropy per baryon varies between 3 and several 10's of $k_{B}$.

BH formation significantly affects the disk properties. If the central object collapses to a BH, approximately half of the disk mass is swallowed inside the apparent horizon within a dynamical timescale, and the maximum density decreases to a few times $10^{12}~{\rm g~cm^{-3}}$. Disk hosting a BH at their center are more compact and achieve higher temperatures and entropies ($\Delta s \simeq 2~{k_B/{\rm baryon}}$ ) than disks hosting a NS remnant \cite{Perego:2019adq}.

Magnetic fields are not expected to play an important role in the inspiral, but they might affect the postmerger evolution \cite{Duez:2006qe, Kiuchi:2017zzg}. Even weak initial fields can be amplified up to values in excess of $10^{16}$ Gauss by a number of mechanism. These include flux freezing and compression, the Kelvin-Helmholtz instability at the collisional interface \cite{ Kiuchi:2015sga}, 
the magneto-rotational instability \cite{Duez:2006qe, Kiuchi:2017zzg}, 
and magnetic field winding \cite{Duez:2006qe}. 

A crucial question is whether or not ordered large-scale fields are formed by dynamo processes after the initial amplification. 
Ordered fields have the potential to power relativistic jets \cite{Bucciantini:2011kx, Ruiz:2016rai}, or drive mildly relativistic outflows \cite{Metzger:2018uni, Fernandez:2018kax}. 
However, even an unstructured magnetic field can generate magnetic stresses. Because the MHD instabilities known to operate in the merger and postmerger operate on length scales as small as few tens of meters to centimeters, it is presently impossible to perform fully-resolved, global binary \ac{NS} merger simulations with realistic initial conditions. Extremely high-resolution simulations can resolve the MHD instabilities after merger, but only if the \acp{NS} are already endowed with magnetar-strengths magnetic fields prior to merger, since this pushes the instabilities to larger scales \cite{Kiuchi:2015sga, Kiuchi:2017zzg}.

Angular momentum transport due to MHD turbulence can be parametrized as an effective $\alpha-$viscosity. Simulations including a physically motivated prescription for viscosity in GR find that the remnant becomes more quickly axisymmetric, possibly reducing the postmerger \ac{GW} emission \cite{Radice:2017zta, Shibata:2017xht}. Angular momentum redistribution in the remnant inside the massive \ac{NS} happens on a timescale \cite{Hotokezaka:2013iia}:
\be
t_{\rm rem} \simeq \alpha^{-1}~R_{\rm rem}^2~\Omega_{\rm rem}~c_s^{-2} 
\simeq 0.56~{\rm s} 
\left( \frac{\alpha}{0.001} \right)^{-1}
\left( \frac{R_{\rm rem}}{15{\rm km}} \right)^2
\left( \frac{\Omega_{\rm rem}}{10^4{\rm kHz}} \right)
\left( \frac{c_s}{0.2c} \right)^{-2} \, ,
\label{eq:trem}
\ee
where $\Omega_{\rm rem}$ and $c_s$ are the remnant angular velocity and typical sound speed, respectively. The removal of the differential rotation leads to a more uniformly rotating object and possibly to its gravitational collapse \cite{Hotokezaka:2013iia}. Inside the Keplerian disk, the angular momentum redistribution causes matter accretion \cite{Fernandez:2015use, Fujibayashi:2017puw, Fernandez:2018kax, Miller:2019dpt} 
on a timescale
\be
t_{\rm disk} \simeq \alpha^{-1} \left(\frac{H}{R} \right)^{-2}
\Omega_{\rm K}^{-1} \simeq 0.78~{\rm s} \left(
\frac{\alpha}{0.02} \right)^{-1} \left( 
\frac{H/R}{1/3}\right)^{-2} 
\left( \frac{M_{\rm rem}}{2.5 \Msun} \right)^{-1/2}
\left( \frac{R_{\rm disk}}{100\,{\rm km}} \right)^{3/2} \, ,
\ee
where $M_{\rm rem}$ is the mass of the central remnant and $R_{\rm disk}$ the radial scale of the disk.

At the end of the GW-dominated phase, neutrino emission becomes the most relevant cooling mechanism \cite{Eichler:1989ve, Rosswog:2003rv, Sekiguchi:2011zd}. 
In particular, hot and dense matter produces neutrinos of all flavors that are eventually emitted to infinity. The typical neutrino mean free path is $\lambda_{\nu} \simeq \left( n_{\rm B} \sigma_0 \left(E_{\nu}/m_e c^2 \right)^2 \right)^{-1} \simeq 24.6~{\rm m} \left( \rho/10^{14}{\rm g~cm^{-3}} \right)^{-1} \left( E_{\nu}/10~{\rm MeV} \right)^{-2}$, where $n_B$ is the baryon density, $\sigma_0 \simeq 4G_F^2 \left(m_e c^2 \right)^2/(\pi (\hbar c)^4)\simeq 1.76 \times 10^{-44}~{\rm cm^2} $ is the typical neutrino cross section scale, and $E_{\nu}$ the neutrino energy. Assuming $T_{\rm rem} \simeq 20~{\rm MeV}$  to be the characteristic temperature of a central remnant that has not (yet) collapsed to a BH, thermal neutrinos' ($E_{\nu} \simeq 3.15 ~T_{\rm rem}$) optical depth is $\tau_{\nu} \simeq R_{\rm rem}/\lambda_{\nu} = \mathcal{O}(10^4)$. Thus, neutrinos are radiated on the diffusion timescale \cite{Perego:2014fma}: 
\be
t_{\rm diff} \simeq \frac{\tau_{\nu} R_{\rm rem}}{c} \simeq 4.28~{\rm s} 
\left( \frac{R_{\rm rem}}{15~{\rm km}} \right)^{-1}
\left( \frac{M_{\rm rem}}{2.5 \Msun} \right)
\left( \frac{T_{\rm rem}}{20~{\rm MeV}} \right)^2 \, .
\label{eq:tdiff}
\ee

Charged current reactions bring neutrinos in thermal and weak equilibrium with matter. As the temperature increases, $\mu_n - \mu_p + \mu_e < 0$ deep inside the remnant and $\bar{\nu}_e$ dominate over $\nu_e$, since the latter are suppressed by degeneracy. Antineutrino abundances are expected to be $Y_{\bar{\nu}_e}\simeq 0.015$, but the impact of trapped neutrinos seems overall marginal \cite{Foucart:2015gaa, Perego:2019adq}. Neutrinos optical depths are much closer to unity inside the disk, so neutrinos with average energies diffuse and stream out within a few milliseconds. These neutrino cooled disks are locally very close to weak equilibrium with $Y_e \simeq 0.1$, because they regulate themselves to a mildly degenerate state ($\mu_e / k_{B}T \sim 1-3$) due to the negative feedback of degeneracy on the cooling rate \cite{Beloborodov:2008nx}. 

The decompression and the heating up of cold matter initially in neutrino-less weak equilibrium leads to its {\it leptonization}, implying $L_{\bar{\nu}_e} \gtrsim L_{{\nu}_e}$, at least during the early post-merger phase. Because free neutrons are abundant, the absorption opacities for $\nu_e$ are larger than those for $\bar{\nu}_e$, while pair processes, responsible for keeping $\nu_{\mu,\tau}$ and their antiparticle in equilibrium, decouple at much larger densities and temperatures inside the remnant, namely $\rho \gtrsim 10^{13}{\rm g~cm^{-3}}$ and $T \gtrsim 8 ~{\rm MeV}$ \cite{Perego:2014fma, Endrizzi:2019trv}. Accordingly, \ac{BNS} simulations including neutrino transport predict the mean neutrino energies at infinity $E_{\nu_e} (\sim 10~{\rm MeV}) \lesssim E_{\bar{\nu}_e} (\sim 15~{\rm MeV}) \lesssim  E_{\nu_{\mu,\tau}} (\sim 20~{\rm MeV})$, with more massive binaries and softer \ac{EOS} resulting in higher mean energies \cite{Sekiguchi:2011zd, Endrizzi:2019trv}.

Due to the strong dependence of the cross-sections on the incoming neutrino energy, neutrinos with different energies decouple from matter from very different regions. While average energy $\nu_e$ and $\bar{\nu}_e$ decouple in the disk at densities between a few and several times $10^{11}{\rm g~cm^{-3}}$, respectively; low energy neutrinos decouple at around $10^{13} {\rm g~cm^{-3}}$ along spheroidal neutrino decoupling surfaces \cite{Perego:2014fma, Endrizzi:2019trv}. The corresponding large variety of relevant thermodynamical conditions implies the need of a coherent treatment of strong and weak interactions over several order of magnitudes in particle densities and temperatures, as well as of an energy dependent treatments of neutrino transport in merger simulations.

The role of neutrino oscillations in \ac{BNS} mergers is largely unexplored. While it is unlikely that neutrino oscillations play a relevant role in the dynamics and fate of the remnant, they might impact the properties of the ejecta by changing the flavor content of the irradiating neutrino fluxes. The fact that electron antineutrino have the largest luminosities and decouple from smaller radii allows for a new kind of oscillation known as matter-neutrino oscillations to occur a few tens of km above the remnant, possibly affecting the properties of the polar ejecta, \eg, Refs.~\cite{Zhu:2016mwa, Tian:2017xbr}. Stability analysis have also shown that neutrino pairs are potentially unstable against fast-flavor conversions immediately above the neutrino decoupling surfaces \cite{Wu:2017drk}. However, only more detailed calculations using the neutrino quantum kinetics equations \cite{Volpe:2015rla,Richers:2019grc} and taking into account the collision integral, as well as the angular and energy distributions of neutrinos emerging from the remnant, will properly address the relevance and the impact of neutrino oscillations.

The \ac{EOS} of NS matter has a clear imprint on the merger dynamics and on the observables. While the low density part of the nuclear \ac{EOS} ($\rho \lesssim \rho_0$) is reasonably well known, large uncertainties still affect the high density part, \eg, Refs.~\cite{Hebeler:2013nza, Oertel:2016bki}. These uncertainties concern both the nature of the nucleonic interaction and the relevant thermodynamics degrees of freedom in ultra-dense environments. In particular, the appearance of new species is expected to decrease the degeneracy of nucleonic matter, lowering the pressure and softening the \ac{EOS}. These particles include hyperons and nucleonic resonances, \eg, Ref.~\cite{Vidana:2010ip}, but also significant fractions of pions and muons due to the high temperatures reached inside the remnant, \eg, Ref.~\cite{Fore:2019wib}. A QCD phase transition to deconfined quark matter is expected to occur at very high densities (and possibly temperatures), but the onset of this transition as well as its type are still largely unconstrained \cite{Busza:2018rrf}.

During the merger, the appearance of these new degrees of freedom can potentially impact the stability of the remnant (\S \ref{sec:merger.outcome}).

\subsection{Fate of the Remnant}
\label{sec:merger.outcome}
The outcome of \ac{BNS} mergers depends on the binary parameters and on the (poorly known) \ac{NS} \ac{EOS}. In particular, whether and when a \ac{BH} forms is primarily determined by the total mass of the binary measured at infinite separation $M = M_A + M_B$, and by the maximum mass supported by the \ac{EOS} for a nonrotating \ac{NS} $\MTOV$ \cite{2016nure.book.....S}. However, finite temperatures and non beta-equilibrated composition effect, as well as the binary mass ratio and the spins, might also affect the merger outcome. 

Sufficiently massive and/or compact binaries form \acp{BH} promptly during merger, \ie, within a dynamical time ($\lesssim 1\ {\rm ms}$). For comparable mass systems, prompt \ac{BH} formation has been empirically determined to occur if $M \gtrsim M_{\rm thr} = k_{\rm thr} \MTOV$, $k_{\rm thr} = 1.3{-}1.7$ being an \ac{EOS} dependent quantity \cite{Shibata:2005ss, Shibata:2006nm, Hotokezaka:2011dh, Bauswein:2013jpa}. An alternative condition is that prompt \ac{BH} formation occurs if $\kappa_2^T \lesssim 43{-}73$ (equivalently if $\tilde\Lambda \lesssim 338{-}386$) \cite{Zappa:2017xba, Agathos:2019sah}. The threshold mass for prompt \ac{BH} formation in unequal mass binaries is not well constrained, but simulations indicate that $M_{\rm thr}$ is smaller for these binaries \cite{Bauswein:2017vtn}. Several works have explored the dependency of $k_{\rm thr}$ on the \ac{EOS} and have shown the existence of \ac{EOS}-insensitive relations linking $k_{\rm thr}$ to the compactness of a reference $1.6 \Msun$ \ac{NS} $C_{1.6}$ predicted by each \ac{EOS}, or to the compactness of the maximum mass nonrotating \ac{NS} $C_{\max}$ \cite{Hotokezaka:2011dh, Bauswein:2013jpa, Koppel:2019pys}. Prompt \ac{BH} mergers are commonly thought to be \ac{EM}-quiet, because in most of these cases all of the matter is engulfed by the \ac{BH} horizon before photons (or even neutrinos) can escape. For this reason, GW170817 is thought not have undergone prompt \ac{BH} formation \cite{Margalit:2017dij, Bauswein:2017vtn}. However, it is important to emphasize that an \ac{EM} counterpart is still possible even with prompt BH formation for binaries the binaries with large mass ratios \cite{Kiuchi:2019lls}.

Binaries not undergoing prompt \ac{BH} formation result in the formation of massive \acp{NS} that are at least temporarily supported against gravitational collapse by the fast rotation \cite{Baumgarte:1999cq, Rosswog:2001fh, Shibata:2005ss, Shibata:2006nm, Sekiguchi:2011zd, Hotokezaka:2013iia, Bernuzzi:2015opx}. 
These remnants are classified as supramassive \acp{NS} (SMNS) \acused{SMNS} if $\MTOV \leq M \leq \MRNS$, $\MRNS$ being the maximum mass predicted by the zero temperature \ac{EOS} for a rigidly rotating \ac{NS}, or hypermassive \acp{NS} (HMNS) \acused{HMNS}, otherwise \cite{Baumgarte:1999cq}. \acp{HMNS} are thought to be supported by differential rotation, while \acp{SMNS} can be supported even after differential rotation has been erased by viscosity. Very low mass systems with $M < \MTOV$, if they exist in Nature, are expected to form stable massive \acp{NS} (MNS) \acused{MNS}. 
It is important to emphasize that this classification is based on properties of equilibrium models and ignores the dynamical nature of the remnant. For example, the fate of the remnant depends not only on its total mass, but also on the amount angular momentum, which in turn is set by the stars radii and spin. Moreover, $\MRNS$ and $\MTOV$ are agnostic to thermal or magnetic effects which can impact the stability of the remnant in nontrivial ways \cite{Kaplan:2013wra, Radice:2018xqa}. The fate of \acp{SMNS} or \acp{HMNS}, especially those with masses close to $\MRNS$, is unclear: some \acp{HMNS} could lose mass due to viscous processes and remain stable over secular timescales and, conversely, some \acp{SMNS} might collapse due to finite temperature effects \cite{Radice:2018xqa}. We call a remnant \emph{short lived} if it collapses during the \ac{GW} dominated phase of the evolution, ${\sim}10{-}20\ {\rm ms}$ of the merger \cite{Bernuzzi:2015opx, Zappa:2017xba}). Otherwise, we call the remnant \emph{long lived}. See Fig.~\ref{fig:overview}.

Remnants that do not collapse on a timescale of a few seconds -- \emph{very long-lived remnants} -- eventually achieve uniform rotation \cite{Radice:2018xqa}. Afterwards, their evolution is driven by the continued emission of \acp{GW} due to residual ellipticity and by \ac{EM} torques, until enough angular momentum is lost to trigger their collapse, or until the stars settle to nonrotating equilibria. The duration of this phase depends on the magnitude of the dipole component of the magnetic field and on the ellipticity of the remnant. The magnetar model for \acp{SGRB} invokes the presence of such very long-lived remnants to explain the X-ray tails seen in about a third of the \acp{SGRB} \cite{Zhang:2000wx, Lasky:2013yaa, Fan:2013cra}. 
Using these models to fit the X-ray tails of \acp{SGRB} provides possible estimates for lifetimes for these remnants which range from tens of seconds to a few hours \cite{Fan:2013cra, Ravi:2014gxa}. 
An important aspect of these models is that the amount of rotational energy that the remnant needs to shed in order to collapse is of the order of a few $10^{52}\ {\rm erg}$. To be consistent with the inferred \ac{EM} energetics of \ac{SGRB} and of GW170817, this energy cannot be primarily radiated in the \ac{EM} channel. Instead, a significant fraction of this energy has to be radiated as \acp{GW} and might be directly detectable for a nearby event \cite{Fan:2013cra} 
It is worth mentioning that there are alternative explanations for the X-ray tails that do not invoke the presence of very long-lived remnants, \eg, \cite{Oganesyan:2019jij}. 

If detected, the \acp{GW} emitted by the merger remnant offer a direct way to observe its fate. However, \ac{GW} searches for the postmerger signal from GW170817 provided only weak upper limits \cite{Abbott:2017dke, Abbott:2018hgk}.

The \ac{EM} data can also be used to constrain the fate of the remnant, although in a model dependent way. As already mentioned, the very presence of an \ac{EM} counterpart disfavours the prompt \ac{BH} formation for the \ac{NS} merger in GW170817 \cite{Margalit:2017dij, Bauswein:2017vtn, Radice:2017lry}. 
Whether or not the merger remnant was hypermassive or supramassive is less clear. The predominant interpretation due to Margalit \& Metzger \cite{Margalit:2017dij} is that GW170817 formed a short-lived remnant. The reason is that, as mentioned above, a long-lived remnant would have injected a few times $10^{52}\ {\rm erg}$ of rotational energy into the ejecta, which can be excluded from observations. Thus Margalit \& Metzger argued that the remnant must have been an \ac{HMNS}, although, as we mentioned above short-lived remnants need not to be necessarily \acp{HMNS}. Because $\MRNS \simeq 1.2 \MTOV$ for most viable \ac{NS} \acp{EOS} \cite{Breu:2016ufb}, assuming that GW170817 was hypermassive implies $\MTOV \lesssim 2.2 \Msun$ \cite{Margalit:2017dij}. 

Other groups have instead interpreted GW170817 in the context of the magnetar model for \acp{SGRB} \cite{Ai:2018jtv, Li:2018hzy, Piro:2018bpl}. To avoid the constraint of Margalit \& Metzger, these models invoke a very long lived remnant (days to months) endowed with a small dipole magnetic field \cite{Ai:2018jtv}. In this way only a modest amount of energy is injected into the outflows. This alternative interpretation would imply a larger maximum mass for nonrotating \acp{NS} $\MTOV \gtrsim 2.2\ \Msun$, and up to an order of magnitude smaller ejecta masses ${\sim}10^{-3}\ \Msun$. The reduced ejecta mass estimate arises because, in these models, the energy injected into the outflows by the central remnant supplements that due to radioactive heating, so a reduced amount of radioactivity is needed to explain the UV/optical/infrared data \cite{Li:2018hzy}. Piro et al.~\cite{Piro:2018bpl} found a sub-threshold ($\gtrsim 3\sigma$) X-ray flare in the Chandra data for GW170817 at about 160 days after the merger that they interpret as evidence of a remnant that has not yet collapsed to \ac{BH}. However, a follow up analysis by Hajela et al.~\cite{Hajela:2019mjy} did not find evidences for X-ray variability.

\subsection{Multimessenger Signatures}
The postmerger phase imprints itself in the many multimessenger signatures of \ac{BNS} mergers. As discussed in more detail in \S \ref{sec:ejecta}, neutron rich material is ejected dynamically as the stars interact \cite{Rosswog:1998hy, Hotokezaka:2012ze, Bauswein:2013yna, Wanajo:2014wha, Radice:2018pdn}, 
and on secular timescales after the merger \cite{Lee:2009uc,Perego:2014fma, Fernandez:2015use, Siegel:2017nub, Fujibayashi:2017puw, Fernandez:2018kax, Miller:2019dpt}. 
This material undergoes r-process nucleosynthesis and synthesizes heavy elements \cite{Eichler:1989ve, Wanajo:2014wha, Cowan:2019pkx}. 
The radioactive decay of by-products of the r-process powers a transient with a quasi-thermal spectrum, the so-called \ac{kN} \cite{Metzger:2019zeh}, 
which was observed in association with GW170817. 

Even before GW170817, it was thought that \ac{NS} mergers could generate ultrarelativistic jets, and that these jets would power gamma-ray flashes (\ac{SGRB}), as well as UV/optical afterglows \cite{Eichler:1989ve, Berger:2013jza}. 
The mechanism for jet launching and the radiative processes responsible for the prompt gamma-ray flash are still debated, while the UV/optical afterglow is known to originate because of the interaction between the jet and the interstellar medium \cite{Kumar:2014upa}. Proposed mechanisms for jet launching include magnetic-field-mediated energy extraction from a remnant spinning \ac{BH} \cite{Blandford:1977ds, Ruiz:2016rai}, 
magnetized winds from a remnant magnetar \cite{Zhang:2000wx, Bucciantini:2011kx}, or neutrino-antineutrino powered fireballs \cite{Eichler:1989ve}. 
In the case of GW170817, a gamma-ray flash was detected by both the INTEGRAL and the Fermi satellites with a delay of 1.7 seconds from the merger \cite{Monitor:2017mdv}. 
Unlike \acp{SGRB} seen at cosmological distance, the \ac{SGRB} in GW170817 was observed off-axis, and is possibly originating from the wings of a structured jet, or from the interaction between the jet and the merger debris \cite{Lazzati:2017zsj, Xie:2018vya}. 

If the remnant avoids gravitational collapse for a sufficient time for all trapped neutrinos to escape, its integrated MeV neutrino luminosity is expected to be a few times $10^{52}\ {\rm erg}$, comparable to that of a regular core-collapse supernova \cite{Radice:2018xqa}. Unfortunately, this translates to a detection range limited to our galaxy for current and even next generation neutrino experiments such as SuperK or DUNE \cite{Palenzuela:2015dqa}. Given that the merger rate in our galaxy is of one merger every $\O(10^4)$ years, the prospects for detection are not rosy. High energy GeV/TeV neutrinos and photons generated by nuclear collisions in the \ac{SGRB} jet, or in the remnant magnetosphere \cite{Murase:2006dr, Fang:2017tla} 
might instead be detectable with detectors like IceCube or VERITAS to distances of tens of Mpc, depending on the lifetime of the remnant, the binary inclination, and the poorly known physical conditions in the jet, so they are a promising possible ``new messenger'' from NS mergers.

\begin{figure}
    \begin{minipage}{0.49\textwidth}
    \includegraphics[width=\textwidth]{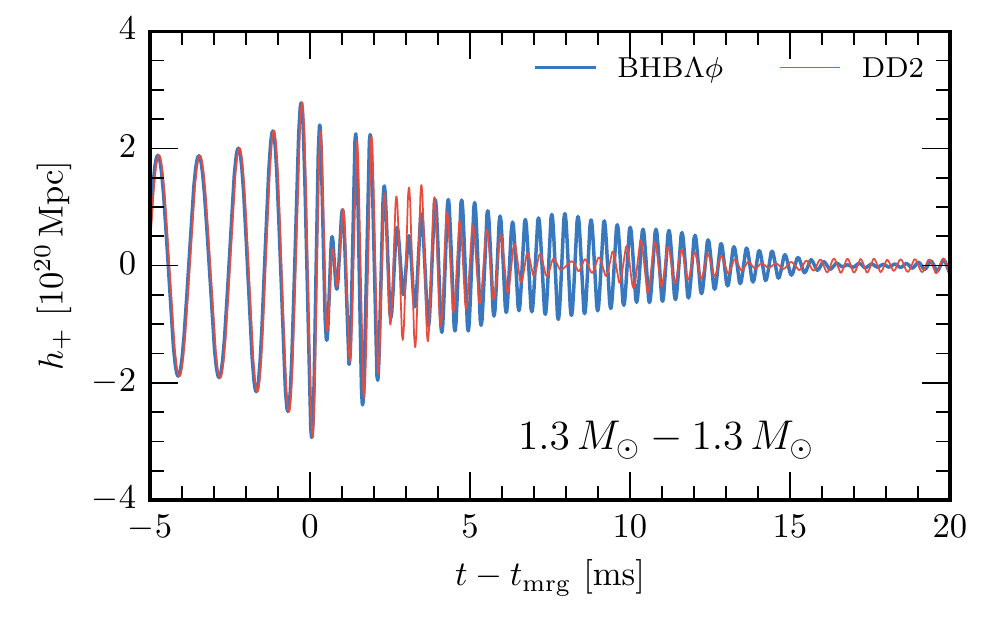}
    \end{minipage}
    \hfill
    \begin{minipage}{0.49\textwidth}
    \includegraphics[width=\textwidth]{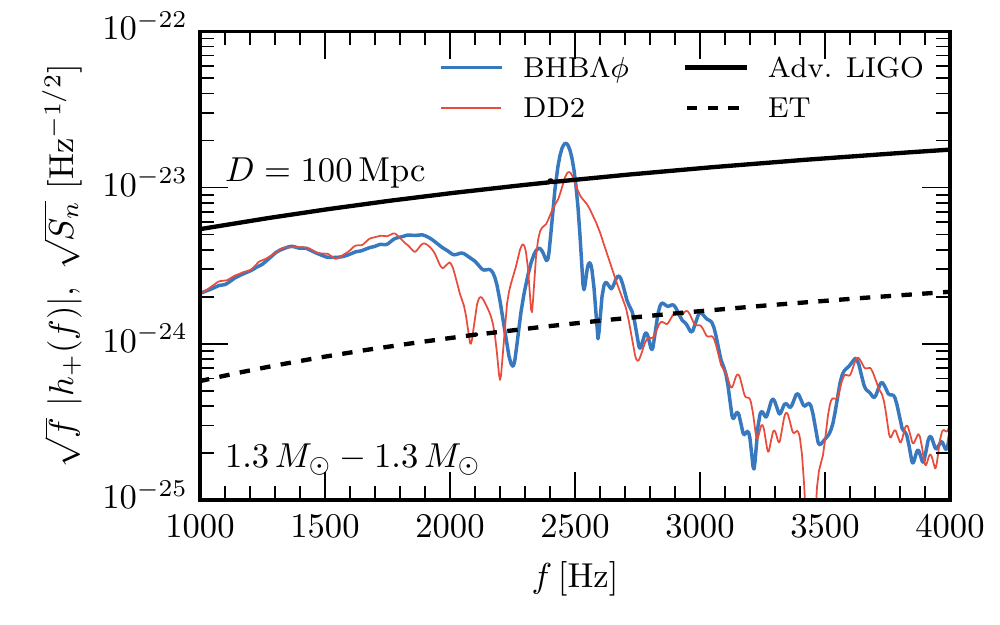}
    \end{minipage}
    \caption{GW strain (left panel) and power spectrum (right panel) for two BNS systems with component masses $1.3\ \Msun$ and $1.3\ \Msun$ simulated with either the DD2 EOS or the BHB$\Lambda\phi$ EOS. The two EOSs are identical at the densities relevant for the inspiral, but they diverge in the postmerger due to the appearance of $\Lambda$ hyperons in the BHB$\Lambda\phi$. See Ref.~\cite{Radice:2016rys} for more details on the simulations.}
    \label{fig:dd2_vs_bhb_fpeak}
\end{figure}

The postmerger is also the phase with the largest \ac{GW} luminosity: up to ${\sim} 0.1\ M_\odot c^2 \simeq 2\times 10^{53}\ {\rm erg}$ are radiated over ${\sim}10{-}20\ {\rm ms}$ \cite{Bernuzzi:2015opx, Zappa:2017xba}. However, most of the \ac{GW} energy is emitted at high-frequency, outside of the main sensitivity band of the detectors \cite{Shibata:2005ss, Bauswein:2011tp, Hotokezaka:2013iia, Rezzolla:2016nxn}, so it will be observable only with 3rd generation detectors, or for very nearby events \cite{Chatziioannou:2017ixj}.  The postmerger \ac{GW} spectrum is characterized by the presence of discrete features or peaks. The main one $f_2$, is a broad peak at frequencies ${\sim}2{-}4\ {\rm kHz}$ \cite{Bauswein:2011tp, Rezzolla:2016nxn, Dietrich:2016hky, Dietrich:2016lyp}. 
For a fixed total mass of the system $f_2$ is found to correlate with $R_{1.6}$ \cite{Bauswein:2011tp, Hotokezaka:2013iia}, with only a weak dependency on mass ratio and \ac{NS} spin \cite{Dietrich:2016hky, Dietrich:2016lyp}. 
A more general quasi-universal relation has been found to link $f_2$ and $\kappa_2^T$ (or $\tilde\Lambda$) in Bernuzzi et al.~\cite{Bernuzzi:2015rla}. Both relations could be used to tightly constrain the \ac{NS} \ac{EOS} if the postmerger signal is detected \cite{Hotokezaka:2013iia, Bernuzzi:2015rla, Chatziioannou:2017ixj, Easter:2018pqy, Tsang:2019esi}. 

Interestingly, even though the densities reached in the postmerger are up to a factor of a few larger than those of the inspiral, these quasi-universal relations imply that the postmerger \ac{GW} signal should be determined by the lower density physics that fixes $R_{1.6}$ and $\kappa^T_2$. The reason for this is that $f_2$ is initially set by the orbital frequency of the stars at merger, which is known to depend only on $\kappa^T_2$ \cite{Bernuzzi:2014owa}. Subsequently, the rate at which angular momentum and binding energy are radiated in \acp{GW} are proportional to each other, so $f_2$ remains roughly constant as the massive \ac{NS} contracts, at least until the last few cycles prior to collapse \cite{Bernuzzi:2015rla, Bernuzzi:2015opx, Dietrich:2016lyp, Radice:2017zta}. This trend is confirmed in simulations that included second order phase transitions after merger. In these simulations the energy liberated by the phase transition boosts the overall \ac{GW} luminosity, but has only a small impact on $f_2$ \cite{Radice:2016rys}. See also Fig.~\ref{fig:dd2_vs_bhb_fpeak}. The only exception is the case in which a strong first order phase transition is present after merger \cite{Bauswein:2018bma}. 
In these cases the merger remnant contracts within a single dynamical timescale, so it is the angular momentum and not the angular velocity -- as was the case for more gradual contraction due to second order phase transitions -- to be approximately conserved. A strong first order phase transition could then be revealed by a tension between the $\kappa_2^T$ inferred from the inspiral signal and that inferred from the postmerger signal \cite{Bauswein:2018bma}, assuming that the phase transition does not result in immediate \ac{BH} formation \cite{Most:2018eaw}.

\section{Matter Ejection, Kilonovae, and Nucleosynthesis}
\label{sec:ejecta}
The ejection of neutron rich material is possibly one of the most important consequence of \ac{NS} mergers \cite{Shibata:2019wef}. The ejecta are thought to undergo the r-process and produce heavy nuclei, making \ac{NS} mergers an important, if not dominant, astrophysical site of production for these elements \cite{Cowan:2019pkx}. The associated \ac{kN} signal was observed in GW170817 and, because of its quasi-isotropic character, it is considered to be the most promising \ac{EM} counterpart for future events \cite{Metzger:2019zeh}.

For the low entropy conditions relevant for \ac{NS} mergers, the outcome of the r-process nucleosynthesis is primarily determined by the electron fraction in the ejecta $Y_e$ \cite{Lippuner:2015gwa}. If $Y_e \lesssim 0.2$, then the ejecta produces second and third r-process peak elements with relative abundances close to Solar. If $Y_e \gtrsim 0.3$, then the material is not sufficiently neutron rich to produce lanthanides. Instead first r-process peak elements are produced. The transition between these two outcomes is for $Y_e \simeq 0.25$ and is very sharp. Not only the nucleosynthesis yield change drastically with $Y_e$, but also the photon opacity in the material changes by orders of magnitude \cite{Tanaka:2013ana, Kasen:2013xka}, drastically altering the timescale and the effective blackbody temperature of the \ac{kN} emission \cite{Metzger:2019zeh}. High $Y_e$ outflows power \acp{kN} peaking in the UV/optical bands within a few hours of the merger (the so-called blue \ac{kN}), while low-$Y_e$ outflows power \acp{kN} peaking in the infrared over a timescale of several days (the so-called red \ac{kN}). In the case of GW170817 both a blue and a red component of the \ac{kN} were observed, suggesting that the outflow had a broad range of compositions, with at least a fraction of the outflow being free of lanthanides.

\begin{figure}
    \begin{minipage}{0.49\textwidth}
        \includegraphics[width=\textwidth]{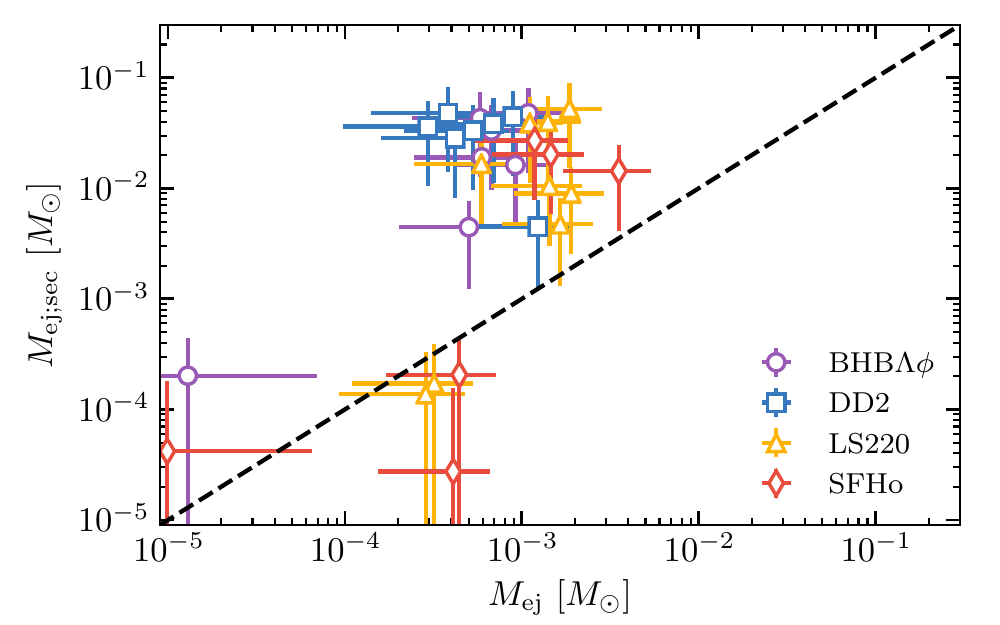}
    \end{minipage}
    \hfill
    \begin{minipage}{0.49\textwidth}
        \includegraphics[width=\textwidth]{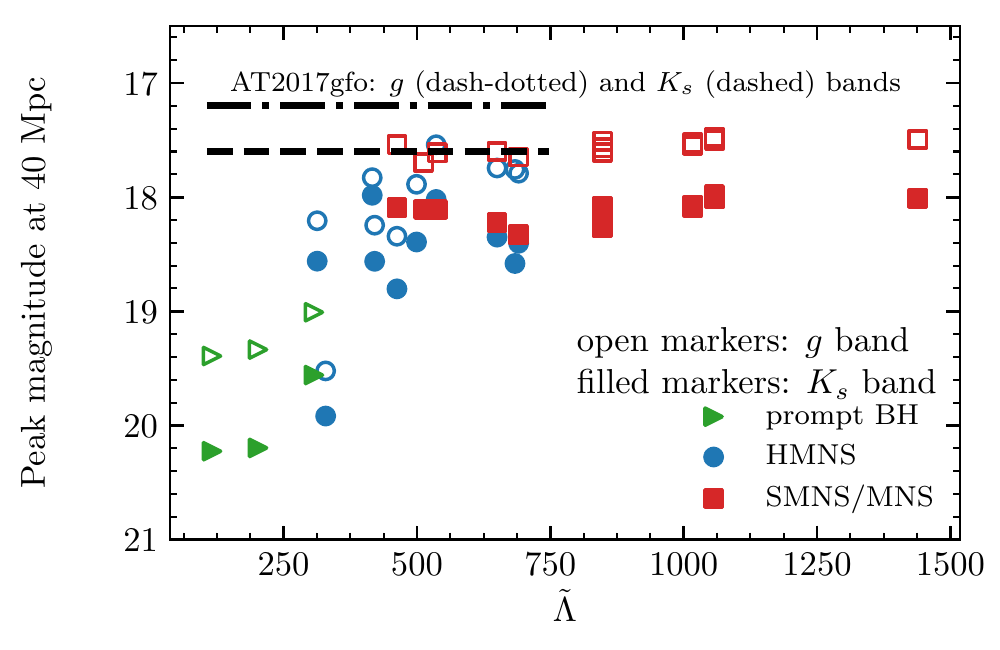}
    \end{minipage}
    \caption{Estimated dynamical and secular ejecta masses (left panel), and light curve peak magnitudes in $g$ and $K_s$ bands (right panel) as a function of the tidal parameter $\tilde\Lambda$ for the BNS models studied in Radice et al.~\cite{Radice:2018pdn}. For comparison we also show in the right panel the magnitudes of AT2017gfo, the \ac{kN} associated with GW170817. The secular ejecta mass is estimated assuming $20\%$ of the remnant disk to become unbound. For most binaries, the secular component of the outflows dominates. In the case of binaries with compact \acp{NS} forming \acp{BH} promptly or shortly after merger, the \ac{kN} peak magnitudes are dimmer due to the smaller amount of ejecta. The peak in the $K_s$ band follows always in time the one in the $g$ band. However, the light curve evolution is much faster and the reddening more significant for smaller $\tilde{\Lambda}$. Note that for the analysis presented in the right panel the light curves are computed for an observer at 30 degrees, while the impact of neutrino irradiation from the remnant, which is expected to enhance these differences even more, is not accounted for. Adapted from Radice, Perego, Hotokezaka, Fromm, Bernuzzi, and Roberts, \textit{Binary Neutron Star Mergers: Mass Ejection, Electromagnetic Counterparts and Nucleosynthesis}, Astrophys.J. 869 (2018) no.2, 130 by permission of the AAS.}
    \label{fig:mass.ejection}
\end{figure}

Part of the outflow is generated on a dynamical timescale: the so-called \emph{dynamical ejecta}. A fraction of this material is ejected due to tidal torques close to the time of merger \cite{Rosswog:1998hy, Radice:2016dwd, Dietrich:2016hky}, especially in the case of very asymmetric binaries. Another fraction is due to shocks generated during and particularly after the merger, when bounces and launches a shock wave into the forming debris cloud \cite{Hotokezaka:2012ze, Bauswein:2013yna, Sekiguchi:2016bjd, Dietrich:2016hky, Radice:2018pdn}.

\ac{GR} merger simulations indicate the mass of the dynamical ejecta ranges from $10^{-4}\ \Msun$ to $10^{-2}\ \Msun$ and that it has characteristic velocities of $0.1{-}0.3\ c$ \cite{Hotokezaka:2012ze, Bauswein:2013yna, Sekiguchi:2016bjd, Radice:2018pdn}. The tidal ejecta is very neutron rich $Y_e \sim 0.1$ and cold, while the shocked ejecta is reprocessed to higher $Y_e$ by pair processes and neutrino irradiation from the central remnant, especially at high latitudes. Indeed, due to the larger equatorial densities, neutrino irradiation is more effective close to the remnant rotational axis. Overall, the dynamical ejecta is found to have a broad range of compositions resulting in a r-process nucleosynthesis pattern close to Solar, with some variations depending on the mass ratio \cite{Sekiguchi:2016bjd, Radice:2018pdn}.

Another component of the outflow, the so-called \emph{secular ejecta}, is due to winds launched after merger \cite{Lee:2009uc, Perego:2014fma, Fernandez:2015use, Siegel:2017nub, Fujibayashi:2017puw, Fernandez:2018kax, Miller:2019dpt, Nedora:2019jhl}. 
In particular, long-term simulations of neutrino-cooled accretion disks around compact objects indicate that $10{-}40\%$ of the remnant disk can be unbound over a timescale of a few seconds. Since \ac{NS} mergers result in the formation of disks with masses up to ${\sim}0.2\ \Msun$, the secular ejecta is thought to constitute the bulk of the outflow. For example, in Fig.~\ref{fig:mass.ejection} we report estimated dynamical and secular ejecta masses from a large collection of \ac{NR} simulations with microphysics reported in Ref.~\cite{Radice:2018pdn}. 

Like the dynamical ejecta, the secular ejecta are also launched by different physical mechanisms. At high latitudes, neutrinos from the remnant and the disk drive a moderately high-$Y_e$ wind \cite{Perego:2014fma, Miller:2019dpt} 
unbinding a few $10^{-3}\ \Msun$ of material. Close to the equatorial plane, viscous effects transport angular momentum causing the disk to spread. Once the accretion rate drops below a critical threshold, neutrino cooling becomes ineffective and the disk thermally expands \cite{Beloborodov:2008nx}. Subsequently, nuclear recombination of nucleons into alpha particles provide sufficient energy to unbind ${\sim}10{-}20\%$ of the disk and produce outflows with characteristic velocities ${\sim}0.1c$ \cite{Lee:2009uc, Fernandez:2015use, Fahlman:2018llv}. 
MHD effects can enhance the outflows masses and asymptotic velocities from these disks, especially at early times, and boost the disk ejection fraction to up to ${\sim}40\%$ \cite{Siegel:2017nub, Fernandez:2018kax}. Finally, \ac{GR} simulations have shown that, if the remnant does not collapse to \ac{BH}, viscous processes in the remnant can drive even more massive and fast outflows \cite{Fujibayashi:2017puw, Radice:2018xqa, Nedora:2019jhl}.

Overall, the composition of the secular ejecta, and hence the properties of the \ac{kN} and the nucleosynthesis yield, are found to depend sensitively on the lifetime of the remnant. For example, in Fig.~\ref{fig:mass.ejection} we show how the peak magnitudes in optical and infrared wavelengths for different \acp{kN} change as a function of the tidal parameter $\tilde\Lambda$ (and hence the merger outcome) for the \ac{BNS} systems studied in \cite{Radice:2018pdn}. We remark that this analysis only accounted for the different remnant disk masses, \ie the differences arise because mergers producing short lived remnants also result in the formation of smaller and more compact disks \cite{Radice:2018xqa, Perego:2019adq}. However, the presence of a long-lived remnant is expected to further affect the \ac{EM} counterpart and enhance the trend shown in the figure by irradiating the ejecta with neutrinos \cite{Fernandez:2015use}. Nevertheless the variability of the \ac{kN} color light curves predicted in these models is very significant. For these reasons, \ac{kN} observations might be promising tools to indirectly probe the outcome of \ac{NS} mergers. Unfortunately, the potential of these observations is hindered by the lack of a quantitative understanding of secular mass ejection in mergers. All of the published postmerger simulations adopted somewhat artificial initial conditions (not derived from merger simulations), or neglected important physical effects such as neutrino emission and absorption, or assumed axisymmetry, or did not follow the evolution for sufficiently long times.

The \ac{kN} associated with GW170817 had both a blue and a red component\footnote{See Ref.~\cite{Waxman:2017sqv} for an alternative interpretation.} \cite{Villar:2017wcc}. The red component has been commonly attributed to a low-$Y_e$ secular ejecta, due to its inferred large mass ${\sim}0.04\ \Msun$ and low velocity ${\sim}0.1 c$. The origins of the blue \ac{kN} are less clear. Simple light curve fitting and spectroscopy suggest that it might have been powered by ${\sim}0.02\ \Msun$ of high-$Y_e$ material expanding with a large velocity ${\sim}0.25c$. Some authors argued that the blue \ac{kN} might have been powered by dynamical ejecta reprocessed to high $Y_e$ by pair processes and neutrino irradiation. 
However, the largest dynamical ejecta mass reported in \ac{GR} simulations is only ${\sim}0.01\ \Msun$ \cite{Sekiguchi:2016bjd}. Moreover, only a small fraction of the dynamical ejecta has a sufficiently large $Y_e$ to power a blue \ac{kN}. More sophisticated multidimensional \ac{kN} models require somewhat smaller ejecta masses to explain the blue \ac{kN} in GW170817, but are still in some tension with simulations \cite{Perego:2017wtu, Kawaguchi:2018ptg}. Alternative explanations of the blue \ac{kN} invoke magnetic effects prior to or after the merger \cite{Metzger:2018uni, Fernandez:2018kax, Radice:2018ghv}, or spiral waves launched in the accretion disk by a long-lived massive \ac{NS} remnant \cite{Nedora:2019jhl}. Future observations of the UV emission from \ac{NS} mergers in the first few hours from the merger might provide additional clues as to which, if any, of these scenarios is correct \cite{Metzger:2018uni}.

While most of the ejecta have velocities $\lesssim 0.4c$, a small fraction of the dynamical ejecta can achieve mildly relativistic velocities up to ${\sim}0.8c$ \cite{Metzger:2014yda, Hotokezaka:2018gmo, Radice:2018pdn, Radice:2018ghv}. This is material that is accelerated as the bounce shock breaks out of the merger debris cloud \cite{Radice:2018pdn}. Some of this material, ${\sim}10^{-6}{-}10^{-5}\ \Msun$ expands sufficiently rapidly to prevent neutrons from capturing on seed nuclei, so this ejecta will undergo free neutron decay and produce an UV bump in the lightcruve on a timescale of one hour \cite{Metzger:2014yda}. As this fast tail of the ejecta interacts with the interstellar medium it is also expected to generate a radio synchrotron remnant, visible on a timescale of months to years after the merger \cite{Nakar:2011cw, Hotokezaka:2018gmo}. In the case of GW170817, the current radio to X-ray synchrotron emission is consistent with the signal from the deceleration of the \ac{SGRB} jet \cite{Hajela:2019mjy}. However, as the \ac{SGRB} afterglows decays, it is expected that the ejecta signal might manifest itself as a bump in the radio light curve. The detection of such signal would confirm that GW170817 produced a massive \ac{NS}, because the acceleration of the outflow to the mildly relativistic velocity required for the radio emission necessitates the shock produced during the merger bounce \cite{Radice:2018pdn}.

\section{Summary Points and Future Issues}
\label{sec:conclusions}
\begin{summary}[SUMMARY POINTS]
\begin{enumerate}
    \item The dynamics of merging NSs is encoded in the GW signal that is thus the primary observable for source identification. The low frequency signal ($\lesssim 50\ {\rm Hz}$) corresponds to the quasi-adiabatic motion and encodes the chirp mass, which was precisely measured for GW170817. Mass ratio and tidal parameters, which need sensitivity at high frequency and precise tidal templates to be well measured, are more uncertain. The early postmerger dynamics has a characteristic transient signal at kiloHertz frequencies that can be computed with \ac{NR} simulations. While full spectrum models are becoming available both merger and postmerger were not observed in GW170817.
    \item \ac{GW} signals from inspiraling \acp{NS} can be used to constrain the \ac{EOS} of matter at up to $\rho \sim 2 \rho_0$. The most robust constrain available to date from GW170817 is that the tidal parameter $\tilde\Lambda$ was smaller than $800$ at 90\% confidence level. More precise constraints on the deformability of the stars, the \ac{EOS}, and the radii of \acp{NS} are available, but are to some extent model and prior dependent.
    \item The postmerger phase probes even higher densities and temperatures of tens to a hundred MeV. However, postmerger \acp{GW} are expected to be emitted predominantly at frequencies $2{-}4\ {\rm kHz}$, outside of the sensitivity band of current GW observatories. The non detection of a postmerger for GW170817 is not constraining for any realistic postmerger model.
    \item High-mass \ac{BNS} mergers result in prompt \ac{BH} formation. Lower mass systems form massive \acp{NS} at least temporarily supported by centrifugal forces against collapse, or even stable \acp{NS}. These different outcomes are imprinted in the characteristics of the \ac{EM} counterparts (or lack thereof). In the case of GW170817, most models favor the formation of a remnant surviving for up to a second and disfavor prompt \ac{BH} formation or very long lived remnants. Alternative outcomes are not completely ruled out: prompt \ac{BH} formation might still be compatible with the observations if the mass ratio of the binary was sufficiently large, and the presence of a long-lived remnant might have been hidden if its dipolar magnetic field was sufficiently weak.
    \item \ac{BNS} mergers produce multimessenger signals in addition to \acp{GW}, including neutrinos and EM radiation over a broad spectrum of energies. The detection of GRB170817A, a \ac{SGRB} associated with GW170817, confirmed that compact mergers are central engines of \acp{SGRB}. The jet launching mechanism and its relation with the merger dynamics are still debated.
    \item Neutron rich ejecta from \ac{NS} mergers synthesize r-process elements and power bright \ac{EM} transients known as kilonovae. \ac{BNS} mergers eject matter on different timescales (dynamical and secular) and through different mechanisms (tidal torques, shocks, nuclear recombination, etc.). The kilonova observed in GW170817 confirms this overall picture and suggests that \ac{BNS} mergers are an important site of production for r-process elements. However, the origin of the outflows in GW170817 are still debated, especially in connection with the UV/optical ``blue'' component of the kilonova that was detected in the first day from the merger. The presence of such a component testifies on the importance of weak reactions in setting the composition of the ejecta.
\end{enumerate}
\end{summary}

\begin{issues}[FUTURE ISSUES]
\begin{enumerate}
    \item How can high-precision measurements of individuals stars masses, spins and tidal parameters be made? As GW observatories become more sensitive and more \ac{NS} mergers are detected, waveform systematic effects will dominate over statistical uncertainties.

    High-fidelity inspiral waveform models capturing the internal dynamics of the stars are needed, but require both analytical improvements and higher quality and longer \ac{NR} simulations than those presently available.
    \item What are the relevant thermodynamical degrees of freedom for the description of matter in merging \acp{BNS}? The formation of muons and pions, and possibly of hyperons, as well as the appearance of QCD phase transitions and their observable consequences need to be studied extensively in simulations. On the one hand, the development of new microphysical \ac{EOS} frameworks and the calculation of the associated weak interaction rates are required. On the other hand, high-resolution \ac{BNS} merger simulations with spectral neutrino-transport are needed to quantify these effects.
    \item What was the origin of the blue component of the \ac{kN} in GW170817? Do \ac{NS} mergers produce all three r-process peak elements? Observations show that the outflows from GW170817 must have been sufficiently neutron rich to produce lanthanides, but there are no direct evidences for the production of higher atomic mass number elements, such as gold. Different models have been proposed to explain the \ac{EM} observations which predict different merger outcomes and nucleosynthesis yields. Early epoch observations of future events might help to distinguish between these possibilities. Ultimately, ab-initio simulations are needed to provide context and to constrain the models used to interpret the \ac{EM} counterpart on the basis of the information provided by the \ac{GW} data.
    \item What was the fate of the \ac{BNS} progenitor to GW170817? There are no self-consistent simulations including inspiral, merger, and postmerger evolution that span all relevant timescales and that include all physical processes known to be important. Extant studies suggest that the outcome of the merger is imprinted in the \ac{EM} and \ac{GW} signals. However, because of the lack of quantitative models, it is not presently possible to confidently constrain the merger outcome of GW170817.

    \item How do the secular ejecta properties depend on the binary parameters? Merger simulations indicate that the structure and mass of the postmerger remnant are sensitive to binary parameters and \ac{EOS}. However, extant long-term postmerger simulations only considered a handful of idealized initial conditions. Thus, it is not clear how the diversity of postmerger configurations is reflected in the \ac{EM} counterpart. For example, will secular winds entrain a roughly constant portion of the disk, or will massive disks evolve in a qualitatively different way depending on the disk mass?
    \item What is the impact of neutrino irradiation and neutrino oscillations on EM counterparts and nucleosynthesis? Neutrino matter interactions play a crucial role in determining the composition of the outflows and, hence, the nucleosynthesis yields and the \ac{EM} emissions of \ac{BNS} mergers. However, on the one hand, extant simulations employ crude approximations to neutrino transport, the state of the art being gray moment schemes. On the other hand, neutrino opacities in dense matter at the conditions relevant for mergers are still not known in a systematic way. Finally, the impact of neutrino oscillations on the observables is still unclear.
  \end{enumerate}
\end{issues}

\bibliographystyle{ar-style5}
\bibliography{references}

\acrodef{ADM}{Arnowitt-Deser-Misner}
\acrodef{AMR}{adaptive mesh-refinement}
\acrodef{BH}{black hole}
\acrodef{BBH}{binary black-hole}
\acrodef{BHNS}{black-hole neutron-star}
\acrodef{BNS}{binary neutron star}
\acrodef{CCSN}{core-collapse supernova}
\acrodefplural{CCSN}[CCSNe]{core-collapse supernovae}
\acrodef{CMA}{consistent multi-fluid advection}
\acrodef{CFL}{Courant-Friedrichs-Lewy}
\acrodef{DG}{discontinuous Galerkin}
\acrodef{kN}{kilonova}
\acrodefplural{kN}[kNe]{kilonovae}
\acrodef{HMNS}{hypermassive NS}
\acrodef{EM}{electromagnetic}
\acrodef{ET}{Einstein Telescope}
\acrodef{EOB}{effective-one-body}
\acrodef{EOS}{equation of state}
\acrodef{FF}{fitting factor}
\acrodef{GR}{general-relativity}
\acrodef{GRLES}{general-relativistic large-eddy simulation}
\acrodef{GRHD}{general-relativistic hydrodynamics}
\acrodef{GRMHD}{general-relativistic magnetohydrodynamics}
\acrodef{GW}{gravitational wave}
\acrodef{ILES}{implicit large-eddy simulations}
\acrodef{LIA}{linear interaction analysis}
\acrodef{LES}{large-eddy simulation}
\acrodefplural{LES}[LES]{large-eddy simulations}
\acrodef{MHD}{magnetohydrodynamics}
\acrodef{MRI}{magnetorotational instability}
\acrodef{NR}{numerical relativity}
\acrodef{NS}{neutron star}
\acrodef{PN}{post-Newtonian}
\acrodef{PNS}{protoneutron star}
\acrodef{SASI}{standing accretion shock instability}
\acrodef{SGRB}{short $\gamma$-ray burst}
\acrodef{SMNS}{supramassive NS}
\acrodef{SPH}{smoothed particle hydrodynamics}
\acrodef{SN}{supernova}
\acrodefplural{SN}[SNe]{supernovae}
\acrodef{SNR}{signal-to-noise ratio}

\end{document}